\documentclass[twocolumn]{aastex63}

\hypersetup{linkcolor=blue,citecolor=blue,filecolor=cyan,urlcolor=blue}
\usepackage{CJK}

\usepackage{longtable}
\usepackage{lipsum} % just for dummy text- not needed for a longtable
\usepackage{multirow}

\usepackage{natbib}
\usepackage{graphicx}
\usepackage{bm}
\usepackage{amsmath}
\usepackage{amsfonts}
\usepackage{amssymb}
\usepackage{xcolor}
\usepackage{mathrsfs}
\usepackage[caption=false]{subfig}
\usepackage{CJK}
\usepackage{ulem}
\graphicspath{{./}{figures/}}

\def\arcsec{$^{\prime\prime}$}
\def\um{$\mu \rm m$}
\def\Mo{$M_{\odot}$}
\def\kms{km~s$^{-1}$}
\newcommand{\CII}{C\textsc{ii}}

\newcommand{\HI}{H\textsc{i}}

\shortauthors{Li et al.}
\graphicspath{{./}{figures/}}
\begin{document}
%\begin{CJK*}{UTF8}{gbsn}
%\title{Template \aastex Article with Examples: v6.3\footnote{Released on June, 10th, 2019}}
\title{The ALMA Survey of 70 $\mu \rm m$ Dark High-mass Clumps in Early Stages (ASHES). II: Molecular Outflows in the Extreme Early Stages of Protocluster Formation}

\correspondingauthor{Shanghuo Li}
\email{shanghuo.li@gmail.com; jzwang@shao.ac.cn}

\author[0000-0003-1275-5251]{Shanghuo Li}
\affil{Shanghai Astronomical Observatory, Chinese Academy of Sciences, 80 Nandan Road, Shanghai 200030, China}
\affiliation{Center for Astrophysics $|$ Harvard \& Smithsonian, 60 Garden Street, Cambridge, MA 02138, USA}
\affiliation{University of Chinese Academy of Sciences, 19A Yuquanlu, Beijing 100049, China}
\affiliation{Korea Astronomy and Space Science Institute, 776 Daedeokdae-ro, Yuseong-gu, Daejeon 34055, Republic of Korea}
% \author{ others}

\author{Patricio Sanhueza}
\affiliation{National Astronomical Observatory of Japan, National Institutes of Natural Sciences, 2-21-1 Osawa, Mitaka, Tokyo 181-8588, Japan}
\affiliation{Department of Astronomical Science, SOKENDAI (The Graduate University for Advanced Studies), 2-21-1 Osawa, Mitaka, Tokyo 181-8588, Japan}

\author{Qizhou Zhang}
\affiliation{Center for Astrophysics $|$ Harvard \& Smithsonian, 60 Garden Street, Cambridge, MA 02138, USA}

\author{Fumitaka Nakamura}\affil{National Astronomical Observatory of Japan, National Institutes of Natural Sciences, 2-21-1 Osawa, Mitaka, Tokyo 181-8588, Japan}
\affiliation{Department of Astronomical Science,  SOKENDAI (The Graduate University for Advanced Studies), 2-21-1 Osawa, Mitaka, Tokyo 181-8588, Japan}

\author{Xing Lu}\affil{National Astronomical Observatory of Japan, National Institutes of Natural Sciences, 2-21-1 Osawa, Mitaka, Tokyo 181-8588, Japan}

\author{Junzhi Wang}
\affil{Shanghai Astronomical Observatory, Chinese Academy of Sciences, 80 Nandan Road, Shanghai 200030, China}

\author{Tie Liu}
\affil{Shanghai Astronomical Observatory, Chinese Academy of Sciences, 80 Nandan Road, Shanghai 200030, China}

\author{Ken'ichi Tatematsu}\affil{National Astronomical Observatory of Japan, National Institutes of Natural Sciences, 2-21-1 Osawa, Mitaka, Tokyo 181-8588, Japan}

\author{James M. Jackson}\affil{SOFIA Science Center, USRA, NASA Ames Research Center, Moffett Field CA 94045, USA}

\author{Andrea Silva}\affil{National Astronomical Observatory of Japan, National Institutes of Natural Sciences, 2-21-1 Osawa, Mitaka, Tokyo 181-8588, Japan}

\author{Andr\'es E. Guzm\'an}\affil{National Astronomical Observatory of Japan, National Institutes of Natural Sciences, 2-21-1 Osawa, Mitaka, Tokyo 181-8588, Japan}

\author{Takeshi Sakai}\affil{Graduate School of Informatics and Engineering, The University of Electro-Communications, Chofu, Tokyo 182-8585, Japan.}

\author{Natsuko Izumi}\affil{National Astronomical Observatory of Japan, National Institutes of Natural Sciences, 2-21-1 Osawa, Mitaka, Tokyo 181-8588, Japan}
\affil{College of Science, Ibaraki University, 2-1-1 Bunkyo, Mito, Ibaraki 310-8512, Japan}

\author{Daniel Tafoya}
\affil{Department of Space, Earth and Environment, Chalmers University of Technology, Onsala Space Observatory, 439~92 Onsala, Sweden}

\author{Fei Li}
\affil{Shanghai Astronomical Observatory, Chinese Academy of Sciences, 80 Nandan Road, Shanghai 200030, China}
\affiliation{University of Chinese Academy of Sciences, 19A Yuquanlu, Beijing 100049, China}

\author{Yanett Contreras}\affil{Leiden Observatory, Leiden University, PO Box 9513, NL-2300 RA Leiden, the Netherlands}

\author{Kaho Morii}\affil{Department of Astronomy, Graduate School of Science, The University of Tokyo, 2-21-1, Osawa, Mitaka, Tokyo 181-0015, Japan}

\author{Kee-Tae Kim}
\affil{Korea Astronomy and Space Science Institute, 776 Daedeokdae-ro, Yuseong-gu, Daejeon 34055, Republic of Korea}
\affil{University of Science and Technology, 217 Gajeong-ro, Yuseong-gu, Daejeon 34113, Republic of Korea}

%\author{}
%\affiliation{}



\begin{abstract}
We present a study of outflows at extremely early 
stages of high-mass star formation obtained  from the ALMA Survey of 
70 \um\ dark High-mass clumps in Early Stages (ASHES). Twelve 
massive 3.6$-$70 \um\ dark prestellar clump candidates were observed 
with the Atacama Large Millimeter/submillimeter Array (ALMA) in Band 6.  
Forty-three outflows are identified toward 41 out of 301 dense cores 
using the CO and SiO emission lines, yielding a detection rate of 14\%. 
We discover  6 episodic molecular outflows associated with low- to 
high-mass cores, indicating that episodic outflows (and therefore 
episodic accretion) begin at extremely early stages of protostellar 
evolution for a range of core masses. 
The time span between consecutive ejection events is much smaller 
than those found in more evolved stages, which indicates that the 
ejection episodicity  timescale is likely not constant over time. 
The estimated outflow dynamical timescale appears to increase with 
core masses, which likely indicates that more massive cores have longer 
accretion timescales than less massive cores. 
The lower accretion rates in these 70 \um\ dark objects compared to 
the more evolved protostars indicate that the accretion rates increase 
with time. 
The total outflow energy rate is smaller than the turbulent energy 
dissipation rate, which suggests that outflow induced turbulence 
cannot sustain the internal clump turbulence at the current epoch.  
We often detect thermal SiO emission  within 
these 70 \um\ dark clumps that is unrelated to CO outflows. 
This SiO emission could be produced by collisions, 
intersection flows, undetected protostars, or other motions.

\end{abstract}



\section{Introduction} 
\label{sec:intro}

Stars influence the dynamics of their natal clouds through a variety of 
feedback by injecting mechanical and thermal/non-thermal  energy 
back to the cloud
\citep{2007prpl.conf..245A,2014prpl.conf..451F}.  The star formation 
feedback may affect the subsequent star formation, for instance, 
through jet and outflow activities, injecting  mass, angular 
momentum, and kinetic energy into the surrounding environment  \citep[e.g.,][]{2010ApJ...709...27W,2014ApJ...790..128F,
2014prpl.conf..451F,2016ARA&A..54..491B}.
Protostellar outflows can efficiently transfer excess angular 
momentum from accretion disks to outer radii, which enables sustained 
accretion from the envelope and/or core to the disk that feeds the 
accreting protostar \citep{1987ARA&A..25...23S,2009ApJ...691..823H,
2009Sci...323..754K,2014prpl.conf..451F,2016ApJ...832...40K}. 
On the other hand, the high momentum outflow can sweep up the 
material in the envelope, diminishing  the potential mass reservoir 
of protostars \citep{2010ApJ...722.1556K,2011ApJ...732...20K}. 
The outflow induced turbulence can help to replenish 
the turbulence that counteracts the  
gravitational collapse/fragmentation \citep{2006ApJ...640L.187L,
2017ApJ...847..104O,2018NatAs...2..896O}, which reduces the overall 
star formation efficiency.

Molecular outflows have been intensively studied in both low-  
and high-mass star-forming regions using different molecular tracers, 
e.g., CO, SiO, HCN, HCO$^{+}$, HNCO, CS, and so on 
\citep{2001ApJ...552L.167Z,2002A&A...383..892B,2005ApJ...625..864Z,
2006ApJ...643..978K,2007prpl.conf..245A,2008ApJ...685.1005Q,
2009ApJ...696...66Q,2010ApJ...715...18S,2012ApJ...756...60S,
2010MNRAS.406..187J,2011ApJ...740L...5C,2011ApJ...735...64W,
2013ApJ...775L..31S,2013A&A...554A..35L,2014ApJ...783...29D,
2014A&A...570A...1D,2015ApJ...802....6S,2016A&ARv..24....6B,
2016A&A...586A.149C,
2017MNRAS.466..248L,2019ApJ...878...29L,2019ApJ...886..130L,
2019ApJ...871..141Q,2019A&A...622A..54P,2020ApJ...890...44B,
2020A&A...636A..38N}. 
Since outflows are a natural consequence of accretion, studying the 
evolution of outflows is critical for understanding the accretion 
history, especially for the extremely early evolutionary stages of 
star-forming regions \citep[e.g.,][]{2019ApJ...886..130L}.
This not only allows to reveal the 
accretion process over time, but also to constrain the initial conditions 
of high-mass star formation. 
Protostars have been observationally shown to have highly variable 
gas accretion rather than constant during formation in both 
low- and high-mass stars 
\citep{2009ApJ...702L..66Q,2014prpl.conf..387A,2014ApJS..211....3P,
2017NatPh..13..276C,2017ApJ...837L..29H,2018A&A...613A..18V}.  
Typical burst intervals from weeks up to 50 kyr have been found in light 
curves of young stars \citep{2013MNRAS.430.2910S,2014ApJS..211....3P,
2018AJ....156...71C}, which could be driven by instabilities in the 
circumstellar disk \citep{2009ApJ...694.1045Z,2010ApJ...719.1896V},  
dynamical interactions between inner objects  
\citep[e.g., disk, companion and planet;][]{2004MNRAS.353..841L},
or inflowing gas streams within the larger scale circumstellar disk 
\citep{2019Sci...366...90A}.

Infall motions are frequently used to study accretion 
processes in the relatively evolved stages of high-mass star 
formation \citep[e.g.,][]{1997ApJ...488..241Z,1988ApJ...324..920K,
2006Natur.443..427B,2016A&A...585A.149W}, 
while it is difficult to directly measure the accretion process during
the earliest phase of high-mass star formation, 
when the stellar embryos 
continue to accumulate the mass during this phase via 
accretion \citep[e.g.,][]{2018ApJ...861...14C}.
Since outflows are accretion-driven \citep{1983ApJ...274..677P,
1986ApJ...301..571P,1991ApJ...370L..31S,2007prpl.conf..261S},
they can be used to study the accretion process. 
Even though molecular outflows only provide indirect evidence of 
accretion, outflows are easily detectable and offer a fossil record of the accretion history of protostars.  
However, a systematic study of the outflows toward the extremely early 
evolutionary phases (e.g., 70 \um\ dark clumps) of star formation with a 
large sample remains scarce due to the weak emission in such   
sources \citep{2019ApJ...886..130L}.

To investigate high-mass star formation in the cluster mode, we have 
carried out high-angular resolution observations towards 12 massive 
70 \um\ dark clumps using the Atacama Large Millimeter/submillimeter Array (ALMA). The sample was selected by combining the ATLASGAL survey \citep{2009A&A...504..415S,2013A&A...549A..45C} and a series of studies from the MALT90 survey \citep{2011ApJS..197...25F,2013PASA...30...38F,2013PASA...30...57J,2015ApJ...815..130G,2016PASA...33...30R,2017MNRAS.466..340C,2017AJ....154..140W}. The source selection is described in detail in \cite{2019ApJ...886..102S}. 

In this work, we focus on identifying molecular outflows and  studying 
their properties in extremely early evolutionary stages of high-mass 
star formation using two molecular lines, CO $J=2-1$ and SiO $J=5-4$. 
Clump fragmentation, based on the dust continuum emission, 
is presented in \cite{2019ApJ...886..102S}. 
Core dynamics will be presented in a companion paper 
(Y. Contreras et al. 2020, in preparation). 

The paper is organized as follows. The observations are described in 
Section \ref{sec:obs}. In Section \ref{sec:res}, we describe the 
results and analysis. We discuss the identified outflows in detail 
in Section \ref{sec:dis}. Finally, we summarize the conclusions in 
Section \ref{sec:con}.

%-------------------------------------------------- 
\begin{figure*}[ht!]
\epsscale{1.18}
\plotone{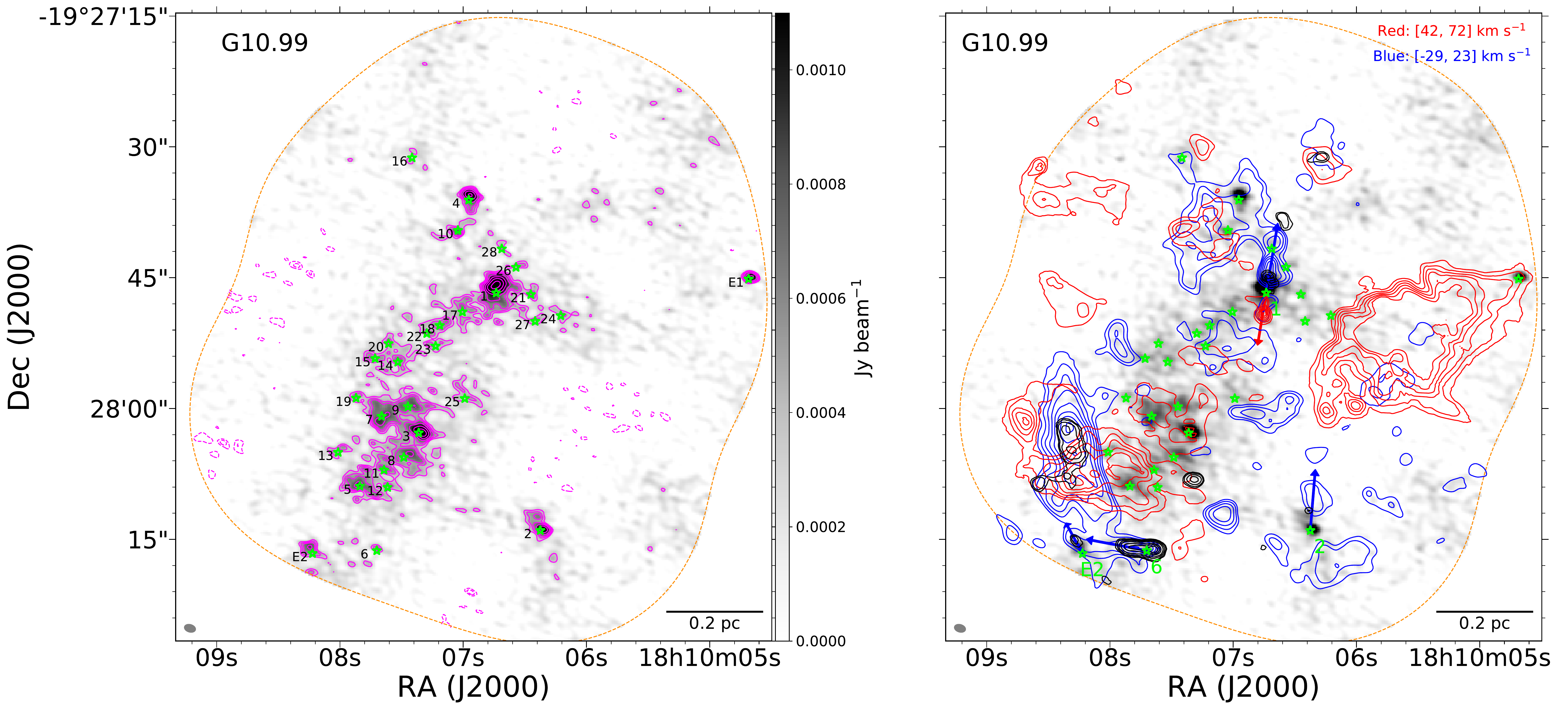}
\caption{ALMA 1.3~mm continuum, CO and SiO velocity integrated 
intensities images for G10.99. The gray images in both panels 
show the 1.3 mm continuum emission.
Dashed orange contour shows 20\% of the sensitivity level of the 
mosaic in the ALMA continuum image.  
The green stars indicate the identified dense cores. 
Left: ALMA 1.3~mm continuum image. The contour 
levels are (-4, -3, 3, 4, 5, 7, 10, 14, 20)$\times \sigma$, with 
$\sigma = $0.115 mJy beam$^{-1}$.  
Right: The blue-shifted (blue contours) and red-shifted (red contours) 
components of CO emission, and SiO velocity-integrated intensity 
(black contours, integration range is between 7 and 39 \kms)  
overlaid on the continuum emission.  The blue and 
red contours levels are (4, 6, 9, 13, 18, 24, 31)$\times \sigma$, 
where the $\sigma$ are 0.05 Jy beam$^{-1}$ km s$^{-1}$ and 
0.03  Jy beam$^{-1}$ km s$^{-1}$ for blue and red contours, respectively.  
The black contours are (3, 4, 5, 6, 7, 8, 9)$\times \sigma$,
with $\sigma =$ 0.02 Jy beam$^{-1}$ km s$^{-1}$. 
On the left side panel, black numbers correspond to the name 
of all cores. On the right side panel, the green numbers 
indicate only the cores associated with molecular outflows.
The beam size of continuum image is shown on the bottom left 
of each panel. The complete figure set (7 images, see 
Figure~\ref{fig:outfmap2}) is available in the online journal.
}
\label{fig:outfmap}
\end{figure*}

%---------------------
\begin{figure*}[ht!]
\epsscale{1.2}
\plotone{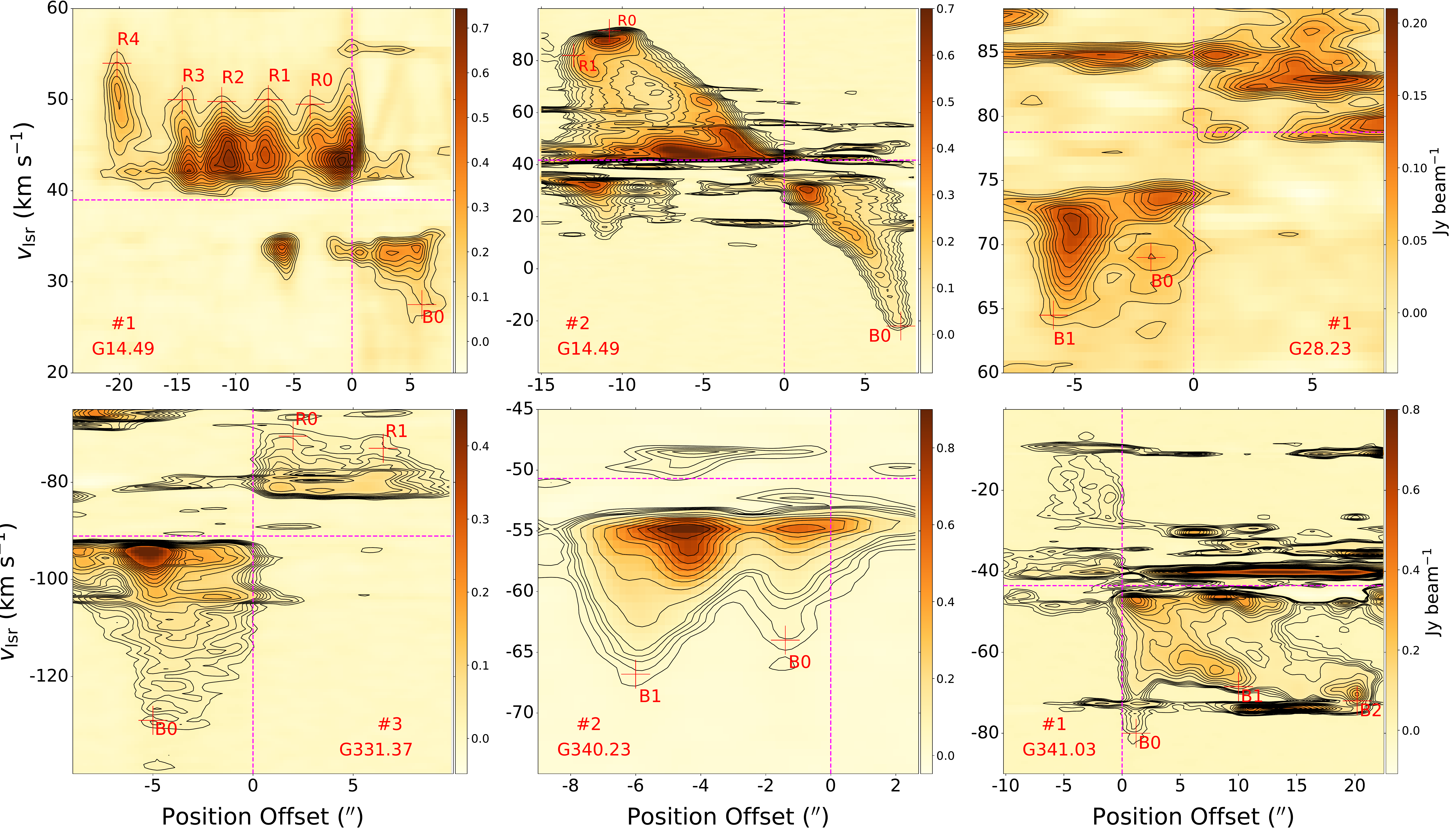}
\caption{Position-velocity (PV) diagram of the CO emission. 
The cut of the PV diagrams is along the CO outflows that are 
indicated by the arrows in Figure~\ref{fig:outfmap}. 
The name of the associated core and clump is displayed  
in each panel. The red crosses mark the locations of the detected 
knots. The vertical and horizontal dashed fuchsia lines  represent the 
position and velocity of dense core, respectively.   
The contour levels are (-3, 3 to 30 by 2 steps)$\times \sigma$, 
with $\sigma = $40 mJy beam$^{-1}$, for \#1--G14.49. 
The contour levels are (-5, 5, 7, 10, 14, 19, 25, 32, 40 to 100 by 
10 steps)$\times \sigma$, with $\sigma = $5 mJy beam$^{-1}$, 
for \#2--G14.49.  
The contour levels are (-5, 5 to 23 by 2 steps)$\times \sigma$, 
with $\sigma = $6 mJy beam$^{-1}$, for \#1--G28.23.  
The contour levels are (-5, 5 to 21 by 2 steps, and 25 to 
80 by 5 steps)$\times \sigma$, with $\sigma = $4 mJy beam$^{-1}$, 
for \#3--G331.37.  
The contour levels are ( -5, 5 to 20 by 3 steps, and 30 to 180 by 20 
steps)$\times \sigma$, with $\sigma = $4.5 mJy beam$^{-1}$, 
for \#2--G340.23.  
The contour levels are (-5, 5 to 20 by 3 steps, and 30 to 150 by 
10 steps)$\times \sigma$, with $\sigma = $3 mJy beam$^{-1}$, 
for \#1--G341.03.  The cut of the \#1--G341.03 PV diagram is along 
the northwest-southeast CO outflow. 
\label{fig:pvdiag}}
\end{figure*}
%-------------------------------------------------- 

\section{Observations} 
\label{sec:obs}
Observations of the 12 infrared dark clouds (IRDCs) were 
performed with ALMA 
in Band 6 (224 GHz; 1.34 mm) using the main 12 m array, 
the Atacama Compact 7 m Array (ACA; Morita Array), and the 
total power (TP; Project ID: 2015.1.01539.S; PI: Sanhueza).

Some sources were observed in different configurations   
\citep[see Table 2 in][]{2019ApJ...886..102S}. 
The uv-taper was used in those sources 
to obtain a similar synthesized beam size of $~$1\farcs2 for 
each source \citep[see also][]{2019ApJ...886..102S}.
The primary beam sizes are 25\farcs2 and 44\farcs6 for 
the 12 m array and ACA at the center frequency of 224 GHz, 
respectively. We made two Nyquist-sampled mosaics with the 
12 m array (10-pointings) and ACA (3-pointings), covering about 
0.97 arcmin$^{2}$ within the 20\% power point, except for the 
IRDC G028.273$-$00.167 that was observed with 11 and 
5-pointings. These mosaics were designed to cover a 
significant portion  of the clumps, as defined by 
single-dish continuum images.  
All sources were observed with the same correlator setup. 

Data calibration was performed using the CASA software 
package versions 4.5.3, 4.6, and 4.7, while imaging was carried 
out using CASA 5.4 \citep{2007ASPC..376..127M}. 
Continuum images were obtained by averaging line-free 
channels over the observed spectral windows with a Briggs's 
robust weighting of 0.5 of the visibilities. 
This achieved an average 1$\sigma$ root mean square (rms) noise 
level of $\sim$0.1 mJy beam$^{-1}$. The beam size  for each 
clump is  summarized in Table~\ref{tab:clumps}. 
For detected molecular lines (N$_{2}$D$^{+}$ 3-2,  DCN 3-2, 
DCO$^{+}$ 3-2, CCD 3-2, $^{13}$CS 5-4, SiO 5-4, C$^{18}$O 
2-1, CO 2-1, and H$_{2}$CO 3-2, CH$_{3}$OH 4-3), we 
used a Briggs's robust weighting of 2.0 (natural weighting) for imaging. 
Using the feathering technique, the 12m and 7m array molecular line data 
was combined with the TP data in order to recover the missing flux. 
This yields a 1$\sigma$ rms noise level of $\sim$0.06 K per channel of 
0.17 \kms\ for the first six lines and $\sim$0.02 K per 
channel of 1.3 \kms\ for the last four lines.

\section{Results and Analysis} 
\label{sec:res}
Using the CO 2-1 and SiO 5-4 lines, we searched for 
molecular outflows toward the 301 embedded dense cores
that have been revealed in the 1.3~mm continuum images of 
the 12 IRDC clumps. \cite{2019ApJ...886..102S} focused on 
294 cores, excluding 7 cores located at the edge of the 
images (20\%-30\% power point). We have included these 
7 cores in our analysis because some of them are associated 
with molecular outflows.  CO was used as the primary 
outflow tracer, while SiO was used as a secondary outflow 
tracer. Figure~\ref{fig:outfmap} shows an overview of the 
identified molecular outflows toward one clump.  
All images shown in the paper use the 12 m array data only 
and prior to primary beam correction, while all measured 
fluxes are derived from the combined data (including 12 m, 7 m, and TP) 
and corrected for the primary beam attenuation.

\subsection{Outflow Detection} 
\label{sec:detection}
For the identification of outflows, we use the 12 m array data only. 
{We follow the procedure described below to identify CO outflows:    
(1) Using 
{\tt DS9}\footnote{\url{https://sites.google.com/cfa.harvard.edu/saoimageds9}},
we inspect the position-position-velocity (PPV) data 
cubes of the CO emission to determine the velocity ranges of the blue- and red-shifted components with respect to the cloud emission.  
(2) These velocity ranges are used to generate the 
velocity integrated intensity maps (moment-0) for both blue- 
and red-shifted components, which in conjunction with the channel 
maps and the CO line profiles are used to determine the direction of CO 
outflows.  
(3) The position-velocity (PV) diagram, which is cut along the 
identified CO outflow, combined with the channel maps and 
line profiles are further used to carefully refine the 
velocity range of each CO outflow to  exclude the 
ambient gas from the cloud. 
However, in some cases (e.g., the outflows associated with core 
\#3/\#5/\#6/\#17/\#32 in the G341.03 clump) 
the outflows cannot be well separated from nearby 
overlapping outflows or ambient gas and those outflows were 
marked as ``Marginal". 
For the outflows, that are unambiguously associated with a dense 
core, and are unambiguously distinguishable from nearby outflows 
and the ambient gas, were marked as ``Definitive". 
Among the 43 identified outflows, 33 are defined as definitive outflows, 
while the remaining are classified as marginal (Table \ref{tab:outflow}).  
The derived physical parameters of the definitive outflows are more 
reliable than those of the marginal outflows (Section~\ref{sec:outfpara}).

The velocity range ($\bigtriangleup v$) of outflows are defined where 
the emission avoid environment gas around the system velocity and 
higher than the 2$\sigma$ noise level, 
where $\sigma$ is the rms noise level in the line-free channels. 
The outflow lobes are defined where 
the corresponding velocity integrated intensity is higher than 
the 4$\sigma$ noise level, where $\sigma$ is the rms of the
integrated intensity map. 
After defining outflows following the 
procedure, we use 
the 12 m, 7 m, and TP combined data to compute outflow parameters. 
Using the CO line, we have identified 
43 molecular outflows toward  41 out of 301 dense cores 
(Table \ref{tab:outflow}), with a detection rate of 14\% (41/301).  
There are 2 cores associated with 2 molecular outflows. 
Among these 41 cores, 28 are associated with  SiO emission 
(detection rate of 9\%). The relatively low detection rate of 
SiO is likely due to the fact that the excitation energy and 
critical density of the SiO 5-4 transition are much higher than 
that of the CO 2-1 transition \citep[e.g.,][]{2019ApJ...886..130L}. 
There are three clumps (G332.96, G340.17 and G340.22) without 
molecular outflows detected in the CO or SiO line.

Sixteen outflows show  bipolar morphology. In addition, we find 
a bipolar outflow without any association with dense cores in 
G331.37 (Figure~\ref{fig:outfmap}), which could be 
due to the associated core mass below our detection limit 
\citep[e.g.,][]{2019A&A...622A..54P}. The remaining  27 outflows 
present a unipolar morphology. The fraction of outflows showing 
unipolar morphology is 63\%, higher than the fraction of 25\% 
and 40\% in W43-MM1 and in a sample of protocluster 
\citep{2020ApJ...890...44B,2020A&A...636A..38N}, respectively. 
Based on the CO emission, we find that the ambient gas is  
frequently found around the dense core. 
Therefore, we might fail to identify some low-velocity and/or 
weak outflows because their emission can be confused with the 
ambient gas. 
We cannot completely rule out the possibility that this bias 
contributes to the high fraction of unipolar outflows.

There are 2 cores, \#2 of G327.11 (\#2--G327.11)
and \#1 of G341.03 (\#1--G341.03), showing two bipolar morphology 
outflows in CO emission (see Figure~\ref{fig:outfmap}). 
The two bipolar outflows are approximately perpendicular to each other 
in both \#2--G327.11 and \#1--G341.03, which suggests that  
these cores encompass multiple protostars with accretion disks 
that are perpendicular to each other.  
Alternatively, these two cores may be showing outflows with wide 
opening angles. However, we favour the first scenario because wide 
angles are expected in much more evolved sources and those expected 
in these IRDCs  that properties are consistent with early stages 
of evolution \citep{2007prpl.conf..245A}.

%---------------------
\begin{figure}[t!]
\epsscale{1.1}
\plotone{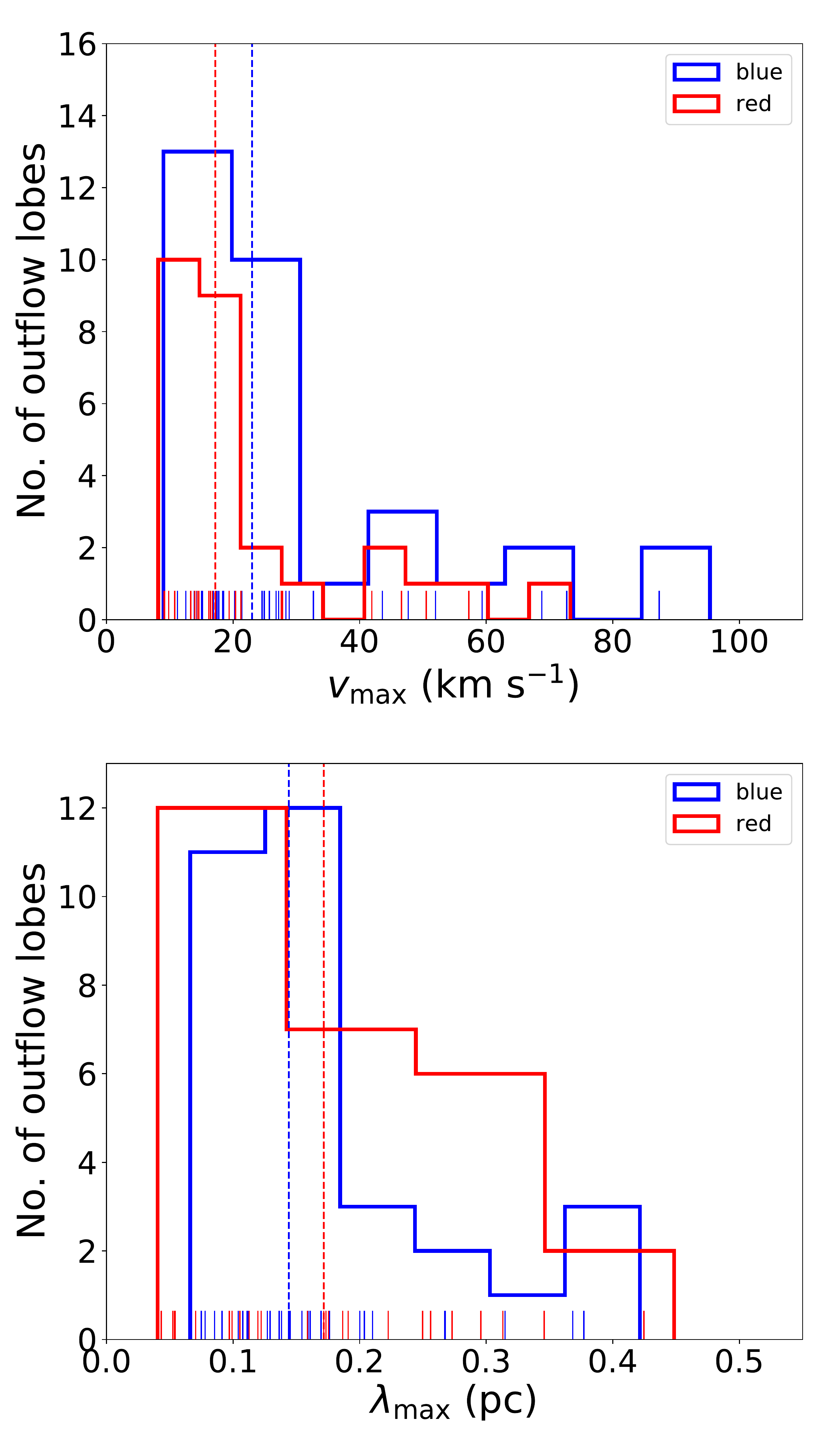}
\caption{Upper and bottom panels show the distributions of the 
maximum velocity ($v_{\rm max}$) and maximum projected distance   
($\lambda_{\rm max}$) of the CO outflow, respectively. 
Red histogram represents the red-shifted gas, while 
blue histogram represents the blue-shifted gas. 
The blue and red dashed lines indicate the median value of 
blue-shifted and red-shifted, respectively. 
\label{fig:vloutfl}}
\end{figure}
%---------------------

%---------------------
\begin{figure*}[ht!]
\epsscale{1.2}
\plotone{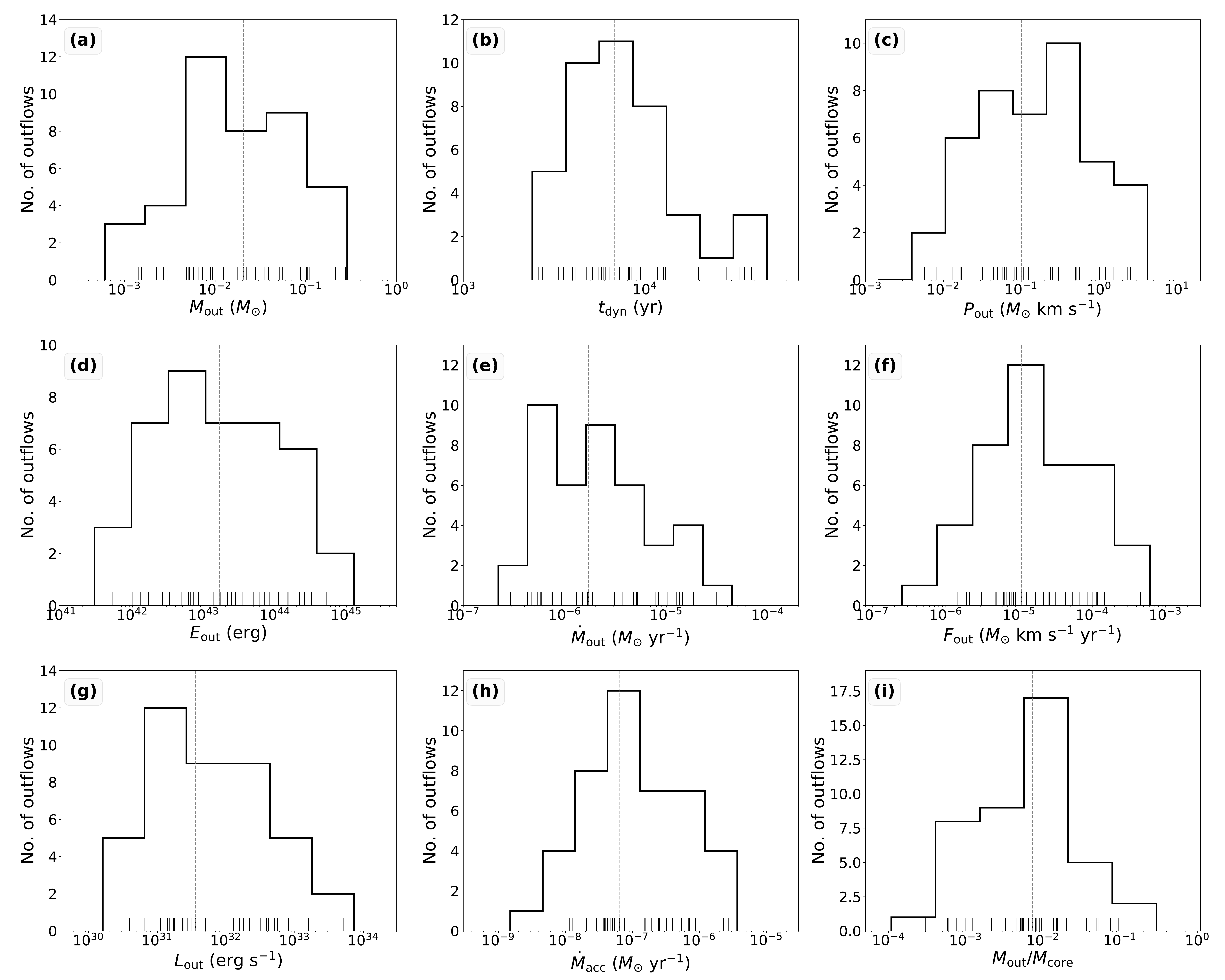}
\caption{Panel (a) to (i) shows the distributions of total outflow 
mass ($M_{\rm out}$), outflow dynamical timescale 
($t_{\rm dyn}$), outflow momentum ($P_{\rm out}$), outflow energy 
($E_{\rm out}$), outflow rate  ($\dot{M}_{\rm out}$),  
mechanical force($F_{\rm out}$), outflow luminosity 
($L_{\rm out}$), accretion rate ($\dot{M}_{\rm acc}$), and 
ratio ($M_{\rm out}/M_{\rm core}$) of  outflow mass to 
core mass, respectively. The black dashed lines 
indicate the median value of each parameter. 
\label{fig:paraoutf}}
\end{figure*}
%---------------------

The velocity integrated intensity is integrated over the entire 
velocity range of the identified red/blue line wings.   
There are several exceptions because some 
foreground and/or background CO emission is found in the velocity 
range of line wings (see Figure \ref{fig:pvdiag}). 
Figure~\ref{fig:outfmap} shows the CO velocity integrated intensity 
map for each clump. The directions of molecular outflows are 
indicated by arrows. 
There are 3 weak CO outflows (\#4--G327.11, \#1--G341.03 and 
\#2--G341.03), identified through PV-diagrams and channel maps.  
The emission from these weak outflows is much smaller than 
the velocity ranges used in the figures (Figure~\ref{fig:outfmap}), 
resulting in a dilution of the signal making the weak outflows 
not visible in the figures.
%and their emission 
%velocity integrated intensities are lower than the 
%contour values of the moment-0 maps shown in Figure~\ref{fig:outfmap}. 

%($v_{\rm max} = |v - v_{\rm lsr}|_{\rm max}$) 

The core systemic velocity is determined by the centroid velocity, 
$v_{\rm lsr}$, of the N$_{2}$D$^{+}$ 3-2 line, while the 
DCO$^{+}$ 3-2 line is used when the N$_{2}$D$^{+}$ line is 
undetected.  The C$^{18}$O 2-1 line is used when both the 
N$_{2}$D$^{+}$ and  DCO$^{+}$ lines are undetected.  
The maximum detected velocities 
(e.g., $v_{\rm max(r)} = {\rm max}(v_{\rm r})$, where 
$v_{\rm r} = |v_{\rm CO} - v_{\rm lsr}|$ 
is a variable representing the red-shifted outflow velocity with 
respect to the core systemic velocity)
of the CO red- and blue-shifted lobes with respect to the core 
systemic velocity range from 9 to up to 95 km s$^{-1}$. 
Figure~\ref{fig:vloutfl} shows the distributions of $v_{\rm max}$ 
for the detected outflow lobes. The mean and median of 
$v_{\rm max}$ are 28 and 21~\kms, respectively. 
There are 9 cores associated with a CO outflow with line wings at 
velocities greater than 30~\kms\ ($v_{\rm max} > 30$) with respect 
to the core systemic velocity. 
The maximum lengths of the CO outflow red-/blue-shifted 
lobes projected on the plane of the sky ($\lambda_{\rm max}$) are in 
the range of 0.04 and 0.45~pc (Figure~\ref{fig:vloutfl}), 
with the mean and median values of 0.25 and 0.23~pc, respectively. 
Note that both $v_{\rm max}$ and $\lambda_{\rm max}$ estimated 
here should be considered as lower limits because the outflow  
emission is limited by the sensitivity of the observations,  
can be confused by nearby outflows and/or line-of-sight emission 
that is not associated with the natal cloud, and these parameters 
have not been corrected for the outflow inclination.

\subsection{Outflow Parameters} 
\label{sec:outfpara}
Since the SiO abundances vary significantly from region to region and 
its emission is much weaker than the CO, we used the CO data to 
compute the outflow parameters  by assuming that the CO line wing 
emission is optically thin. 
Following the approach delineated in \cite{2019ApJ...886..130L} 
(see Appendix \ref{app:equation}), we estimated the main physical 
parameters of each outflow, including the mass ($M_{\rm out}$), 
momentum ($P_{\rm out}$), energy ($E_{\rm out}$), dynamical 
timescale ($t_{\rm dyn}$),  outflow rate ($\dot{M}_{\rm out}$), 
outflow energy rate (also known as outflow luminosity, $L_{\rm out}$) 
and mechanical force ($F_{\rm out}$). 
The calculation of these outflow parameters is summarized in 
Appendix \ref{app:equation}.

The outflow masses range from  0.001 to 0.32~\Mo, 
with a mean mass of 0.044~\Mo\  
(Figure~\ref{fig:paraoutf} and Table~\ref{tab:outflow}). 
We compare the outflow mass to the core mass, and find that 
the outflow mass is smaller than 10\% of the core mass 
(Figure~\ref{fig:paraoutf}),  except for  \#13-G341.03 outflows,  
with a mass of 22\% of the core mass. This is owing to the outflow 
of \#13-G341.03 is significantly contaminated by the nearby 
outflow driven by the \#3--G341.03 (see Figure~\ref{fig:outfmap}). 
The estimated outflow dynamical timescales range from 
0.1 $\times \, 10^{4}$ to 3.2 $\times \, 10^{4}$~yr. 
We used the outflow mass and dynamical time to estimate 
the outflow ejection rates, which are between 
2.4 $\times \, 10^{-7}$ and 5.0 $\times \, 10^{-5}$ 
$M_{\odot}$~yr$^{-1}$. 
The outflow momenta range from 0.2 $\times \, 10^{-2}$ to 
3.4 $M_{\odot}$~km~s$^{-1}$. 
The outflow energies are between 0.4 $\times \, 10^{42}$ and 
1.2 $\times \, 10^{45}$ erg. 
The outflow luminosity  can be estimated by dividing the outflow 
energy by the outflow dynamical timescale, 
$L_{\rm out} = E_{\rm out}/t_{\rm dyn}$, which are in the 
range of 0.2 $\times \, 10^{31}$ and 2.1 $\times \, 10^{34}$ 
erg~s$^{-1}$.  
The outflow mechanical force is computed using the outflow 
momentum and dynamical timescale, 
$F_{\rm out} = P_{\rm out}/t_{\rm dyn}$, which is between 
0.3 $\times \, 10^{-6}$ and 9.8 $\times \, 10^{-4}$
$M_{\odot}$~km s$^{-1}$~yr$^{-1}$.

Assuming that: (1) the outflows are driven by protostellar 
winds from accretion disks \citep[e.g.,][]{2003ApJ...599.1196K,
2003ApJ...585..850M}; (2) the wind and molecular gas 
interface is efficiently mixed \citep{2000prpl.conf..867R}; 
(3) the momentum is conserved between the wind and 
the outflow, with no loss of momentum to the surrounding cloud,  
one can estimate the accretion rate, 
$\dot{M}_{\rm acc} = k\, \dot{M}_{\rm w}$, using  the mass-loss 
rate ($\dot{M}_{\rm w}$) of the wind that is inferred 
from the associated outflow mechanical force, 
$\dot{M}_{\rm w} v_{\rm w} = F_{\rm out}$. 
We assume a ratio  of 
$k\, =\, \dot{M}_{\rm acc}/\dot{M}_{\rm w}\, =\, 3$
between  the mass accretion rate  and the mass ejection rate 
\citep{1998ApJ...502L.163T,2000prpl.conf..789S}. 
The wind velocity $v_{\rm w}$ is adopted to be 500 \kms, which 
ranges from a few~\kms\ to over 1000 \kms\ in young stellar 
objects \citep[YSOs;][]{2016ARA&A..54..491B}.
The derived accretion rate ranges from 2.0 $\times\, 10^{-9}$  
to 3.8 $\times\, 10^{-6}$ $M_{\odot}$~yr$^{-1}$, with a mean 
value of  4.9 $\times\, 10^{-7}$ $M_{\odot}$~yr$^{-1}$. 
These values are much smaller than the predicted accretion rate 
(order $10^{-4}-10^{-3}$ $M_{\odot}$~yr$^{-1}$) in some high-mass 
star formation models 
\citep[e.g.,][]{2003ApJ...585..850M,2010ApJ...709...27W}. 
See discussions of accretion rate below in Section~\ref{sec:accretion}.

Since the inclination angle of the outflow axes is not known, 
the outflow parameters and accretion rate have not been 
corrected for their inclination for all of outflows. The estimated 
outflow parameters of each outflow are summarized in 
Table~\ref{tab:outflow} and Figure~\ref{fig:paraoutf}.

\subsection{Outflow-driven turbulence} 
\label{sec:outftur}
Stars generate vigorous outflows during their formation process. 
These outflows transfer not only mass but also energy into 
their parent molecular cloud.   
Therefore, outflow feedback is considered one of the mechanisms  
to replenish the internal turbulence \citep{2006ApJ...640L.187L,
2010MNRAS.409.1412G,2010ApJ...715.1170A,2011ApJ...737...56N,
2014prpl.conf..243K,2014ApJ...783..115N,2015ApJS..219...20L,
2017ApJ...847..104O}.
To assess whether the outflow-induced turbulence can play a 
significant role in replenishing the internal turbulence for the 
extremely early phase of high-mass star formation, we 
estimate the ratio of the outflow energy rate ($L_{\rm out}$) to 
turbulent energy dissipation rate ($\dot{E}_{\rm turb}$),  
$r_{L} = L_{\rm out}/\dot{E}_{\rm turb}$. 

The turbulent energy ($E_{\rm turb}$) is given by 
\begin{equation}
\label{Etur}
E_{\rm turb} = \frac{1}{2} M (\sqrt{3}\; \sigma_{\rm los, 1D})^{2},
\end{equation}
where $\sigma_{\rm los, 1D}$ is the one-dimensional velocity 
dispersion along the line of sight, which is 
\begin{equation}
\label{sigma1d}
\sigma_{\rm los, 1D} =\frac{\sqrt{\triangle v_{\rm FWHM}^{2} - \triangle v_{\rm chan}^{2}}}{2\sqrt{2\rm \;ln\;2}},
\end{equation}
Here, the $\triangle v_{\rm chan}$ and $\triangle v_{\rm FWHM}$ 
are the channel width and the full width at the half maximum 
(FWHM) of the line emission, respectively. 
Using the physical properties of the clumps (Table \ref{tab:clumps}), 
we obtained the turbulent energies ranging from 
5.4 $\times$ 10$^{45}$ to 4.4 $\times$ 10$^{47}$ erg 
(Table~\ref{tab:impact}).  

The turbulent dissipation rate ($\dot{E}_{\rm turb}$) can be 
calculated as \citep{2007ARA&A..45..565M}
\begin{equation}
\label{dissrate}
\dot{E}_{\rm turb} = \frac{E_{\rm turb}}{t_{\rm diss}},
\end{equation}
where dissipation time, $t_{\rm diss}$, has two different definitions 
based on numerical simulations. 
In the first approach, the turbulent dissipation time of the cloud 
is given by  \citep[e.g.,][]{2007ARA&A..45..565M}
\begin{equation}
\label{Tdis0}
t_{\rm diss} \sim 0.5 \frac{R_{\rm cl}}{\sigma_{\rm 1D}},
\end{equation}
where $R_{\rm cl}$ is the cloud radius. 
The obtained turbulent dissipation times are between 
5.8 $\times$ 10$^{4}$ and 3.6 $\times$ 10$^{5}$~yr, 
yielding turbulent dissipation rates of 
$(0.5-114.7) \times\, 10^{33}$~erg~s$^{-1}$ (Table~\ref{tab:impact}). 
The obtained $t_{\rm diss}$ is comparable to the estimated 
value of a few 10$^{5}$~yr in the Perseus region 
\citep{2010ApJ...715.1170A}, while is smaller than the reported 
value of 1.6 $\times$ 10$^{7}$ yr in the Taurus molecular cloud  
\citep{2015ApJS..219...20L}. The $r_{L}$ lies in the range of 
0.004--4.05, with a median value of 0.08 (Table \ref{tab:impact}). 
The range of values is much lower than those found in relatively evolved star forming-regions, for instance, 0.8$-$12.4 in Perseus,  
2$-$7 in $\rho$ Ophiuchi,  0.6$-$3 in Serpens South  
\citep{2010ApJ...715.1170A,2011ApJ...726...46N,
2015ApJ...803...22P}.

%---------------------
\begin{figure*}[ht!]
\epsscale{1.2}
\plotone{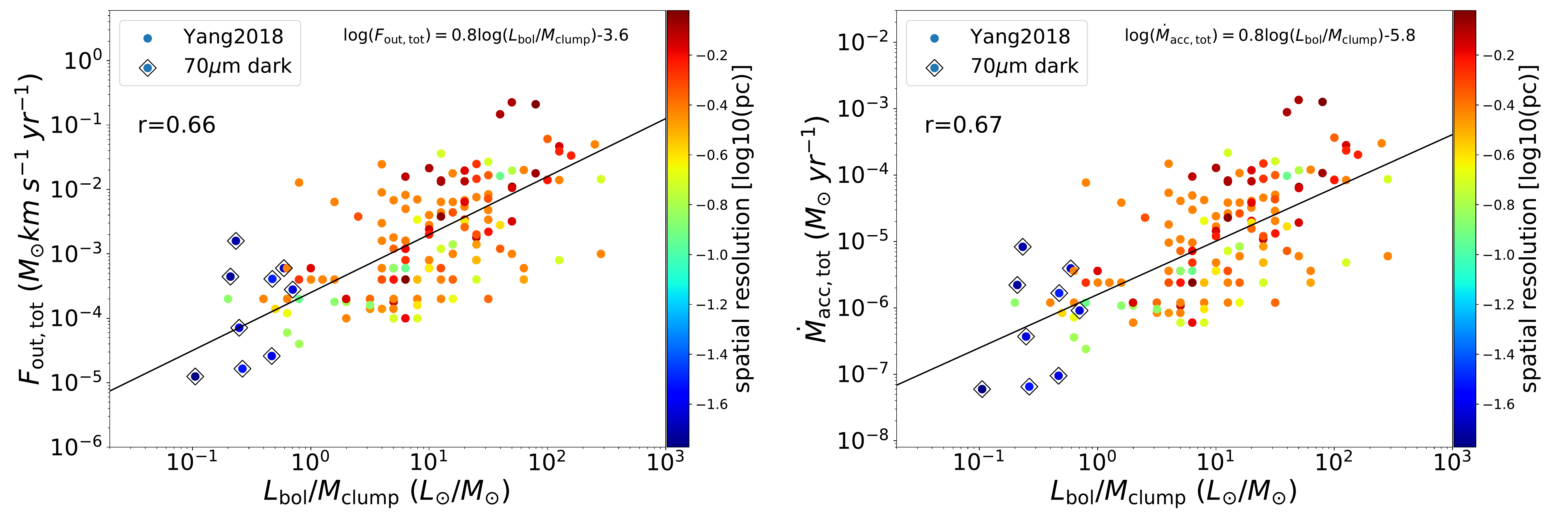}
\caption{Left: plot of the $L_{\rm bol}/M_{\rm clump}$ against the 
$F_{\rm out, tot}$.  The correlation coefficient is 0.66 between 
$L_{\rm bol}/M_{\rm clump}$ and $F_{\rm out, tot}$ in the 
Spearman-rank correlation test. 
The solid black line is the  least-squares fitting result 
${\rm log}(F_{\rm out, tot}) = 0.8{\rm log}(L_{\rm bol}/M_{\rm clump})-3.6$ \citep[see also][]{2018ApJS..235....3Y}.
The colorbar indicates the spatial resolution. 
Right: plot of the $L_{\rm bol}/M_{\rm clump}$ versus the 
$\dot{M}_{\rm acc, tot}$. There is a strong correlation between  
$L_{\rm bol}/M_{\rm clump}$ and $\dot{M}_{\rm acc, tot}$ with a correlation 
coefficient of 0.67. The solid black line is the best least-squares fitting result 
${\rm log}(\dot{M}_{\rm acc, tot}) = 0.8{\rm log}(L_{\rm bol}/M_{\rm clump})-5.8$.
The colorbar indicates the spatial resolution. 
\label{fig:LF}}
\end{figure*}
%---------------------

In the second approach, the turbulent dissipation time can be 
defined as  \citep[e.g.,][]{1999ApJ...524..169M}
\begin{equation}
\label{Tdis1}
t_{\rm diss} \sim  3.9 \frac{\kappa}{M_{\rm rms}}  t_{\rm ff},
\end{equation}
where $M_{\rm rms} = v_{\rm los, 3D}/c_{\rm s} = 
\sqrt{3} v_{\rm los, 1D}/c_{\rm s}$ is the Mach number 
of the turbulence, $t_{\rm ff} = \sqrt{3 \pi / 32 G \rho}$ is the 
free-fall timescale, $\kappa = \lambda_{d}/\lambda_{J}$ 
is the ratio of the driving wavelength ($\lambda_{d}$) to the 
Jeans wavelength ($\lambda_{J} = c_{\rm s} (\pi /G\rho)^{1/2}$). 
The average volume density of a clump is calculated by 
$\rho$ = $3M_{\rm cl}/4\pi R_{\rm cl}^{3}$. 
For a continuous outflow, the outflow lobe length can be used 
to approximate the turbulence driving wavelength according 
to numerical simulations 
\citep{2007ApJ...662..395N,2009ApJ...692..816C}. 
The estimated $t_{\rm diss}$ ranges from 1.3 $\times$ 10$^{4}$ 
to 1.5 $\times$ 10$^{5}$ yr, resulting in turbulent dissipation rates of 
$(3.8-109.5) \times\, 10^{33}$~erg~s$^{-1}$ (Table~\ref{tab:impact}).  
Using the derived $\dot{E}_{\rm turb}$, we find that the $r_{L}$ is 
between 0.001 and 0.54, with a median value of 0.02.

\section{Discussion}
\label{sec:dis}
\subsection{Biases on Outflow Parameters} 
\label{biase}
CO is the most frequently used tracer for studying molecular 
outflows driven by protostars. Note that the outflow parameters 
estimated from the CO line emission are only a lower limit 
considering the observational biases:   
(1) CO fails to trace the outflow in low-density regions, 
which can be traced by other low-density gas tracers (e.g.,  
\HI\ 21 cm and \CII\ 157 \um\ lines); 
(2) optical depth effects on the outflow parameters; 
(3) the limited sensitivity of our observations. 
On average, the combined effects of the above mentioned 
biases could lead to a factor of $\sim$10 underestimate in 
the outflow mass and dynamical properties 
\citep{2014ApJ...783...29D}.  
On the other hand, the unknown inclinations of the outflow 
axis with respect to the line of sight can also introduce 
an uncertainty on the  outflow parameters estimations 
\citep{2014ApJ...783...29D,2019ApJ...886..130L}.  
Assuming all orientations have equal probability from parallel 
($\theta = 0^{\circ}$) to perpendicular ($\theta = 90^{\circ}$) 
to the line of sight, one can compute an mean inclination angle 
$\langle \theta \rangle \approx 57^{\circ}.3$.
For the mean inclination angle, the correction factors are 1.9 
for $v_{\rm out}$;  1.2 for $\lambda_{\rm out}$;   0.6 for 
$t_{\rm dyn}$;  1.9 for $P_{\rm out}$;  3.4 for $E_{\rm out}$; 
and 1.7 for $\dot{M}_{\rm out}$ \citep[see,][]{2014ApJ...783...29D,
2019ApJ...886..130L}.

\subsection{Outflow Properties} 
\label{sec:outfprop}
To study the relationship between the outflow properties and their  
associated dense cores, we have compared core masses and outflow parameters, 
including the outflow dynamical timescale, outflow maximum velocity, outflow 
projected distance, outflow mass, outflow momentum, outflow energy, outflow 
rate, outflow luminosity, and mechanical force.  There are no  clear 
correlations between core masses and outflow parameters, with the 
exception of outflow dynamical timescale, which appears to show a 
positive correlation with the core mass. 
In Section~\ref{sec:accretion}, we discuss the accretion rate 
and outflow mechanical force. 
Then, we discuss the relationship between the outflow 
dynamical timescale and the core mass in  Section~\ref{sec:time}. 
The discussion of emission knots  
in the molecular outflows is presented in Section~\ref{sec:episodic}.

\subsubsection{Accretion Rate and Outflow Mechanical Force}
\label{sec:accretion}
Using the CO outflows, the estimated protostellar accretion 
rates are between  2.0 $\times\, 10^{-9}$  to 
3.8 $\times\, 10^{-6}$~$M_{\odot}$~yr$^{-1}$, with a mean 
value of  4.9 $\times\, 10^{-7}$~$M_{\odot}$~yr$^{-1}$. 
These values are comparable to those derived toward other 70~\um\  
and 24~\um\ dark massive cores \citep{2015ApJ...805..171L,
2015ApJ...804..141Z,2019ApJ...886..130L} and some 
low/intermediate-mass cores \citep{2016A&A...587A..17V,
2020ApJ...896...11F}  following the same approach, while they 
are much smaller than the typical value of a few 
$\times$($10^{-5} - 10^{-4}$) $M_{\odot}$~yr$^{-1}$ reported in 
much more evolved massive cores 
\citep[e.g., 24~\um\ bright cores, 4.5~\um\ bright cores and UCHII 
regions,][]{2009ApJ...702L..66Q,2009ApJ...696...66Q,
2010A&A...517A..66L,2013A&A...558A.125D,2017ApJ...849...25L,
2018ApJ...855....9L}. 
This indicates that the accretion rates likely increase with 
the evolution of star formation.

The bolometric luminosity-to-mass ratio ($L/M$) is believed to be a 
indicator of the evolutionary phase of star formation. 
The early evolutionary phase correspond to a low $L/M$ value, while 
later stages correspond to higher values of $L/M$ 
\citep{2008A&A...481..345M}. Figure~\ref{fig:LF} shows the 
luminosity-to-mass ratio ($L_{\rm bol}/M_{\rm clump}$) versus 
the total outflow mechanical forces ($F_{\rm out, tot}$), 
for our 70~\um\ dark clumps, together with high-mass clumps 
from \cite{2018ApJS..235....3Y}. 
In the latter clumps, the observations have an angular 
resolution of 15\arcsec\ \citep{2018ApJS..235....3Y}, which 
should not be able to spatially resolve  the outflows for 
the embedded cores. 
Thus, we added the outflow parameters (i.e., mechanical force 
$F_{\rm out, tot}$ = $\sum F_{\rm out}$ and accretion rate 
$\dot{M}_{\rm acc, tot}$ = $\sum \dot{M}_{\rm acc}$) for each core within  
a clump for our 70~\um\ dark sources.   
The bolometric luminosities of our clumps are estimated with the 
procedure  introduced by \cite{2017MNRAS.466..340C} (see their 
Equation~3).  The $L_{\rm bol}/M_{\rm clump}$ and 
$F_{\rm out, tot}$ shows a strong positive correlation, over nearly $\sim$4 
order of magnitude in $L_{\rm bol}/M_{\rm clump}$ and $\sim$5 order 
of magnitude in $F_{\rm out, tot}$ \citep[see also,][]{2018ApJS..235....3Y}. 
The correlation coefficient is 0.66 with a p-value of $2\, \times\, 10^{-22}$ 
in a Spearman-rank correlation test, which assess monotonic relationships.
This indicates that outflow forces increase with protostellar  
evolution in the high-mass regime. The outflow forces in our clumps 
are relatively lower compared to other high-mass clumps in 
\cite{2018ApJS..235....3Y}, which is likely due to a later evolutionary 
stage as evidenced by the high $L_{\rm bol}/M_{\rm clump}$ value.

As mentioned in Section \ref{sec:outfpara}, we can estimate the 
accretion rate using the outflow force. Here, we adopt a wind 
velocity of 500 \kms\ and a $\dot{M}_{\rm acc}/\dot{M}_{\rm w}$ 
of 3  for all the clumps in \cite{2018ApJS..235....3Y}. 
Figure~\ref{fig:LF} shows a strong positive correlation between 
the total accretion rate ($\dot{M}_{\rm acc, tot}$) and  
$L_{\rm bol}/M_{\rm clump}$. The Spearman-rank correlation test 
returns a correlation coefficient of 0.67 with a p-value of 
$3\, \times\, 10^{-2}$. This strong correlation suggests that 
the accretion rates indeed increase with the evolution of star 
formation.

For the dense cores detected CO outflows, the estimated 
accretion rates (a few $10^{-7}\, M_{\odot}$~yr$^{-1}$) are 
$\sim$3 orders of magnitude lower than the predicted values 
of high-mass star formation models 
\citep{2003ApJ...585..850M,2010ApJ...709...27W,
2016ApJ...832...40K}.  
The computed accretion rates might be increased by a factor of 
$\sim$2 orders of magnitude in the most extreme cases that have 
the highest ratios of accretion to ejection rates, which varies 
from 1.4 to even 100 in different observational and theoretical 
studies \citep[e.g.,][]{2012ApJ...757...65S,2013A&A...551A...5E,
2014prpl.conf..451F,2016ApJ...832...40K}. 
Assuming a constant ratio of the accretion to mass ejection rate, we found that the 
estimated accretion rates of these protostars are indeed 
significantly lower than the predicted values in the 
theoretical studies, even after accounting for the effects 
of opacity and inclination angle.

%---------------------
\begin{figure}[t!]
\epsscale{1.3}
\plotone{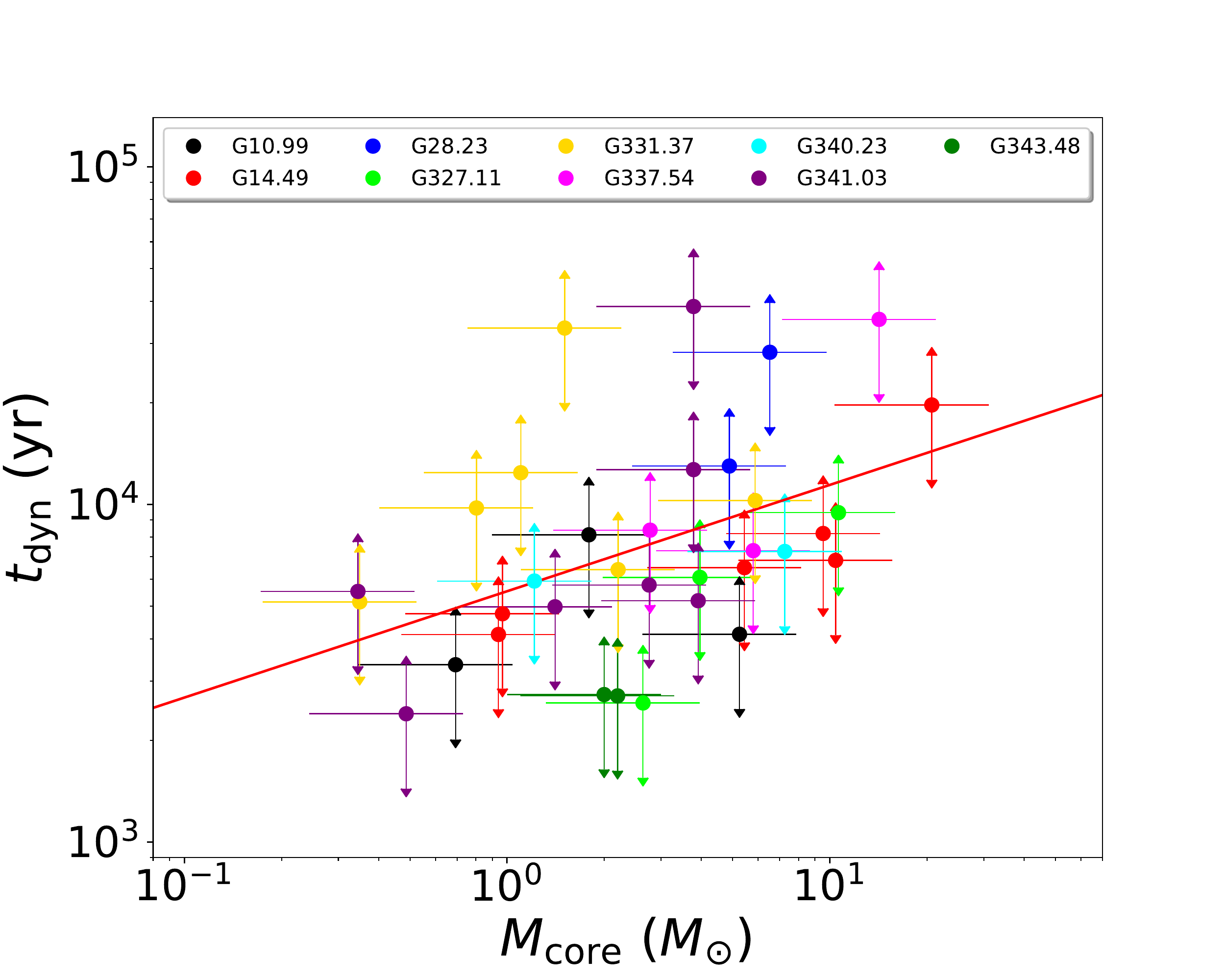}
\caption{Core mass versus the outflow dynamical 
timescale for all cores associated to CO outflows.  
The Spearman-rank correlation test returns a correlation 
with a coefficient of 0.51. 
The solid red line is the best least-squares fitting  result with 
${\rm log}(t_{\rm dyn}) = 0.32{\rm log}(M_{\rm core})+3.74$. 
The error bars of $M_{\rm core}$ correspond to its uncertainty 
of 50\%\ \citep{2017ApJ...841...97S,2019ApJ...886..102S}.  
The error bars of $t_{\rm dyn}$ correspond to  the 
mean inclination angle ($\langle \theta \rangle \approx 57^{\circ}.3$) 
correction. 
\label{fig:MT}}
\end{figure}
%---------------------

\subsubsection{Outflow Dynamical Timescale} 
\label{sec:time}
Figure~\ref{fig:MT} shows the relation between the outflow 
dynamical timescale ($t_{\rm dyn}$) and the core mass 
($M_{\rm core}$). We have removed marginal outflows in this 
plot to eliminate the effect of the ambiguous outflows. 
Consistent with what 
\cite{2019ApJ...886..130L} found for a sample of 70~\um\ dark 
clumps, the outflow dynamical  timescale increases with 
the core mass. 
The correlation coefficient is 0.51 with 
a p-value of 2$\times$10$^{-3}$  in a Spearman-rank correlation 
test. This indicates a moderate correlation between the outflow 
dynamical timescale and the core mass. 
The $t_{\rm dyn} - M_{\rm core}$ relation suggests that more 
massive cores tend to have a longer outflow dynamical timescale 
than less massive cores. 
Since outflows are a natural consequence of accretion, 
the positive relation between  $t_{\rm dyn}$ and $M_{\rm core}$ 
suggests that the more massive cores have a longer accretion 
history than less massive ones in a protocluster. 
Note that the measurement of outflow dynamical timescales is 
affected by observational effects such as the unknown inclination 
angle of the outflow axes. 
The positive correlation between  $t_{\rm dyn}$ and $M_{\rm core}$ 
persists unless the inclination angles of the outflows are close to 
0$^{\circ}$ or 90 $^{\circ}$ (see Figure~\ref{fig:MT}).

\subsubsection{Episodic Ejection} 
\label{sec:episodic}
From Figure~\ref{fig:outfmap}, we find that 6 out of 43 molecular 
outflows  show a series of knots in the integrated intensity map. 
Figure~\ref{fig:pvdiag} presents their PV diagrams along the axis of 
outflows. The Hubble Law and Hubble Wedge features are clearly 
seen in the PV diagram of some outflows 
\citep{1996ApJ...459..638L,2001ApJ...554..132A}. 
Episodic variation of, e.g., mass-loss 
rate and flow velocity, in a protostellar ejection can produce 
internal shocks. This mechanism  has been proposed to produce 
emission knots (which correspond to the Hubble Wedge in the PV 
diagram) in molecular outflows \citep{2001ApJ...554..132A,
2007prpl.conf..245A,2018A&A...613A..18V,2019ApJ...877..112C,
2019MNRAS.483.2563R}.  
Since the accretion is episodic, the outflow as a natural consequence 
of the accretion is episodic as well  
\citep{2015ApJ...800...86K,2016ARA&A..54..491B,
2018A&A...612A.103C}. The Hubble Wedge features are found to be 
independent of the core mass, associated with both low-mass and 
high-mass cores (see Figures~\ref{fig:outfmap} and \ref{fig:pvdiag}). 
This suggests that episodic accretion-driven outflows begin 
in the earliest phase of protostellar evolution for low-mass, 
intermediate-mass, and high-mass protostellar cores in a protocluster.

%---------------------
\begin{figure}[t!]
\epsscale{1.2}
\plotone{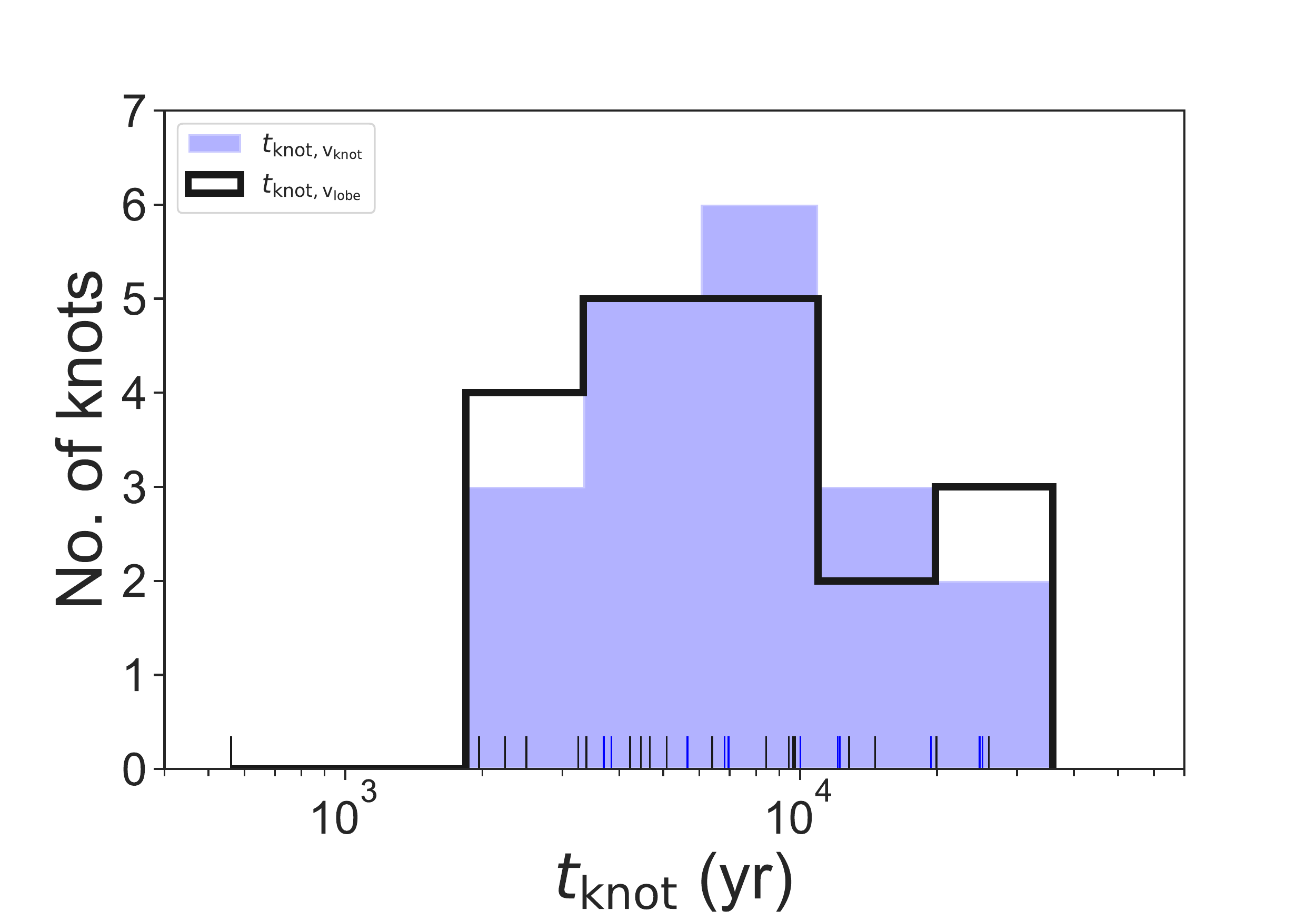}
\caption{The distribution of knots dynamical timescales 
estimated by $v_{\rm knot}$ and $v_{\rm lobe}$. 
\label{fig:tknot}}
\end{figure}
%---------------------

In Section \ref{sec:outfpara} (Appendix~\ref{app:equation}), we 
used the maximum velocity and 
the projected distance of the CO outflows to estimate the outflow 
dynamical timescale, $t_{\rm dyn} = \lambda_{\rm out}/v_{\rm max}$. 
Similarly, we can estimate the dynamical timescale of the identified knots 
using their projected distance ($\lambda_{\rm knot}$) and 
corresponding velocity 
($v_{\rm knot}$) with respect to the associated core, 
$t_{\rm knot} = \lambda_{\rm knot}/v_{\rm knot}$.  From 
Figure~ \ref{fig:pvdiag}, one notes that the knots have different 
velocities. Therefore,  we use $v_{\rm knot}$, instead of the 
mean velocity of detected knots 
($v_{\rm lobe} = \langle v_{\rm knot} \rangle$) as used in 
\cite{2020A&A...636A..38N}, when calculating the knot dynamical 
timescale. 

We visually follow contours in the PV-diagram to define the 
velocity and the projected distance of detected knots (see 
Figure~ \ref{fig:pvdiag}).  The locations of knots in each 
outflow are presented in Figure~ \ref{fig:pvdiag}. 
The mean $v_{\rm knot}$ for all knots is  22 \kms, ranging from 
10 to 60 \kms. The $\lambda_{\rm knot}$ are between 0.02 and 0.39 pc, 
with a mean value of 0.15 pc. 
Uncertainties are conservatively estimated as 2 \kms\ for velocities 
(3 times of channel width), and 0\farcs5 for lengths 
(about half of angular resolution).  
The estimated dynamical timescales of knot range from 
6 $\times\, 10^{2}$ to 2.5 $\times\, 10^{4}$ yr, with the mean 
and median value of 8.6 $\times\, 10^{3}$ and 
6.6 $\times\, 10^{3}$ yr, respectively (see Table~\ref{tab:knots}).  
These values are higher than the value of $100- 6000$ yr found 
for low-mass YSOs 
\citep[e.g.,][]{2009A&A...495..169S,2015Natur.527...70P}, 
but comparable to the value of $400-10^{4}$ yr for high-mass 
YSOs \citep[e.g., W43-MM1;][]{2020A&A...636A..38N} and the 
predicted timescale of a few$\times$($10^{3}  - 10^{4}$) yr 
in the model of \cite{2018A&A...613A..18V}.

To study the time span between consecutive ejection events, 
we compute the timescale difference between knots, 
$\triangle t_{\rm knot}$ = $t_{{\rm knot}, i+1}$ - $t_{{\rm knot}, i}$, 
which ranges from 
3.6 $\times\, 10^{2}$ to 7.2 $\times\, 10^{3}$~yr, with a mean 
value of 4.4 $\times\, 10^{3}$ yr.  The derived   
$\triangle t_{\rm knot}$ is much larger than the reported 
value of $\sim$500~yr for W43-MM1 \citep{2020A&A...636A..38N}, 
even accounting for the typical projection correction and the 
typical measured uncertainty.

By comparing the knots timescales estimated by $v_{\rm knot}$ 
and $v_{\rm lobe}$, we find that the two results agree very 
well with each other (see Figure~\ref{fig:tknot}). 
This suggests that the different knot timescales between 
these studies are not due to the use of two different methods. 
The different  $\triangle t_{\rm knot}$ could be due to the 
different period of episodic ejection between two samples, 
considering that they are significantly different in the 
evolutionary stages.  
Our 70~\um\ dark cores are in a much earlier stage than those 
infrared bright cores in W43-MM1 \citep{2020A&A...636A..38N}. 
If this scenario is correct, the discrepancy likely implies 
that the ejection episodicity timescale is not constant over 
time and accretion bursts may be more often at later stages 
of evolution.

The consecutive Hubble Wedge features seen 
in the PV diagram are not necessarily successive in time for 
the evolved objects in a  very clustered environment. This is 
because they have had enough time to experience numerous 
repeated ejection events.
Therefore, they can create a series in time with  nonconsecutive 
Hubble Wedge features.  In this IRDC study, the detected outflows 
are still in an extremely early phase of protostellar evolution, 
thus the outflows likely trace the early accretion activities. 
This could be another possibility to cause the different 
$\triangle t_{\rm knot}$ between different evolutionary stages 
of star formation.  
The observations of W43-MM1 
by \citet{2020A&A...636A..38N} have a spatial resolution of 
$\sim$2400~AU, which is higher 
than our observations ($\sim$4000--7500 AU). 
Given the different spatial resolutions between the two studies, we 
cannot fully rule out the possibility that the different  
$\triangle t_{\rm knot}$ is due to the effect of
spatial resolutions as higher spatial resolutions 
help to resolve more individual knots. 
All these possibilities  can be tested with the high spatial resolution 
observations of a larger sample of dense cores 
encompassing  different evolutionary stages of star formation.

\subsection{Outflow Impact on the Clumps} 
\label{sec:impact}
The mechanical energy in outflows is found to be comparable 
to the turbulent energy in the more evolved sources of both 
low- and high-mass star formation 
\citep[eg.,][]{2002A&A...383..892B,2005ApJ...625..864Z,
2010ApJ...715.1170A,2014ApJ...783..115N}.  
In addition, the outflow-induced turbulence can sustain the 
internal turbulence \citep{2010ApJ...715.1170A,
2014ApJ...783..115N,2015ApJS..219...20L}. 
In contrast, we find that the total outflow energy 
(2.7 $\times \, 10^{42}\, - \,2.1\, \times \, 10^{45}$ erg) in the 
sample is significantly lower than the turbulent energy 
(3.1 $\times \, 10^{45}\, - \,4.2\, \times \, 10^{47}$ erg).  
This indicates that outflows do not have the energy for 
driving the turbulence in these clumps at the current epoch. 
In addition, the total outflow ejected mass occupies only a tiny 
fraction of the clump mass, with a ratio of 
6 $\times$ 10$^{-6}$ -- 3 $\times$ 10$^{-4}$ 
(see Table~\ref{tab:impact}).

%---------------------
\begin{figure}[t!]
\epsscale{1.3}
\plotone{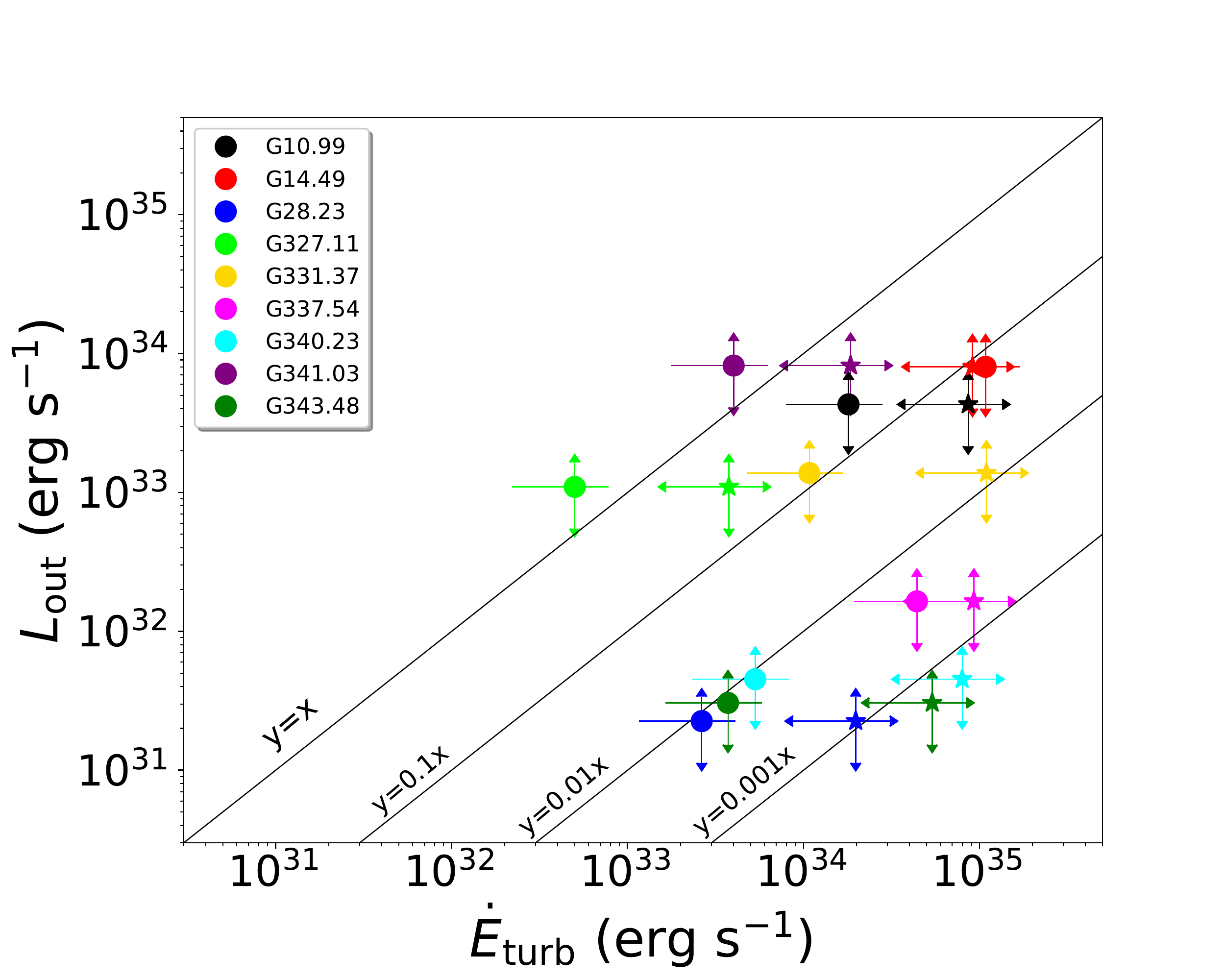}
\caption{The turbulent dissipation rate versus the outflow 
luminosity. The filled circles and filled stars represent the turbulent 
dissipation rate computed by the first  (Equation \ref{Tdis0}) 
and second (Equation \ref{Tdis1})  approaches, respectively.  
\label{fig:LoutEturb}}
\end{figure}
%---------------------

Figure~\ref{fig:LoutEturb} shows the outflow luminosity versus 
the turbulent dissipation rate. 
As mentioned in Section \ref{sec:outftur}, we compute the turbulent 
dissipation time ($t_{\rm diss}$) using two different approaches. 
The determined $t_{\rm diss}$ using the first approach 
(Equation \ref{Tdis0}) is larger than that of the second approach 
(Equation \ref{Tdis1}). The latter is depending on the length of 
outflows, which may introduce uncertainties in the estimation of 
$t_{\rm diss}$.  Overall, the results from both approaches are 
consistent (Figure~\ref{fig:LoutEturb} and Table~\ref{tab:impact}), 
the outflow luminosity is much smaller than the turbulent dissipation 
rate in the majority of clumps, except for G327.11 and G341.03,  
which have relatively larger $r_{\rm L}$.  
The computed $r_{\rm L}$ is smaller as compared to the values 
found in other works for more evolved sources  
\citep[e.g.,][]{2010ApJ...715.1170A,2011ApJ...726...46N,
2015ApJ...803...22P}. 
This indicates that the outflow-induced turbulence produced by these 
early stage cores is not yet severely affecting the internal clump 
turbulence.

%-----------------------------
\begin{figure*}[ht!]
\epsscale{1.2}
\plotone{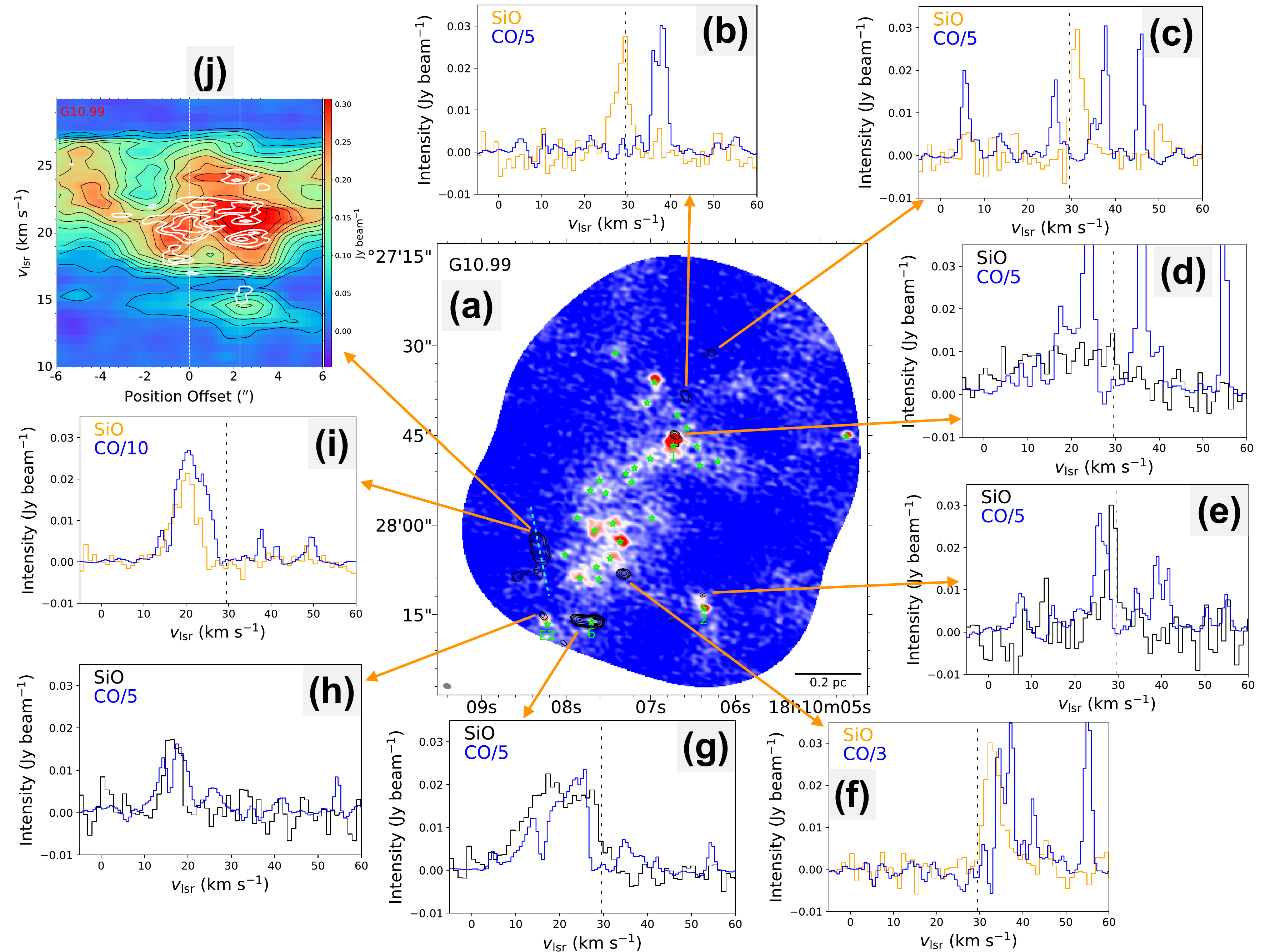}
\caption{Panel a:  SiO velocity-integrated intensity 
(black contours, integration range is between 7 and 39 \kms) 
overlaid on the continuum emission of G10.09. The contours are 
(-4, -3, 3, 4, 5, 6, 7, 8, 9)$\times \sigma$, with 
$\sigma =$ 0.02 Jy beam$^{-1}$ km s$^{-1}$. Green stars indicate 
the dense cores revealed by dust continuum. The continuum beam 
size is plotted at the bottom-left corner. The cyan dashed 
lines indicate the PV cuts path (panel j and k). 
Panel b-i: spectra of SiO 5-4 (orange  and black) and CO 3-2 
(blue) toward several positions across the 
G10.99.  Panel b, c, f and i show the SiO (orange) and CO profiles 
toward the regions where SiO emission appear to neither associated 
with outflows nor dense cores. The black vertical dashed line is 
the clump velocity. 
Panel j: PV diagram of the CO (color-scale and black contours) 
and SiO (white contours) emission across the SiO contours 
(cyan dashed lines).  
The white contours are (3, 4, 5, 6)$\times \sigma$, with  
$\sigma$ = 0.008 Jy beam$^{-1}$. The black contours are 
(7, 14, 21, 28 ...)$\times \sigma$, 
with $\sigma$ = 0.004 Jy beam$^{-1}$. 
The white dashed lines indicate the peak positions of 
SiO contours. 
}
\label{fig:spectSiO}
\end{figure*}
%--------------------------------------

From Table \ref{tab:impact}, we note that the G327.11 has the 
narrowest line width and the lowest mass, resulting in the lowest 
turbulent energy and the lowest turbulent dissipation rate among 
the sample. This seems to be the reason why G327.11 can have the  
outflow-induced turbulence required for replenishing the clump turbulence. 
G341.03 is another exception with sufficient outflow-induced 
turbulence to maintain the clump turbulence. Using the CO line emission, 
we find that it has the highest outflow detection rate (27\%) 
and has the largest outflow luminosity among the whole IRDC sample.  
This implies that outflows can become a major source of 
turbulence in the clumps when the majority of the embedded cores  
evolve to launch strong molecular outflows.

Strong outflows generated by young stars have a disruptive 
impact on their natal clouds. 
Similarly, the outflow energy ($E_{\rm out}$) can be compared to the 
gravitational binding energy of clouds, 
$E_{\rm grav} = GM_{\rm cl}^{2}/R_{\rm cl}$, for evaluating 
the potential disruptive effects of outflows on their clouds.  
The gravitational energy of clouds ranges from 7.4 $\times\, 10^{46}$ 
to 5.3 $\times\, 10^{48}$ erg, which is much larger than the total 
outflow energy for each clump. The  ratio of $E_{\rm out}/E_{\rm grav}$ 
is between 2 $\times\, 10^{-5}$ and 8 $\times\, 10^{-3}$, which 
indicates that the disruptive impact of these outflows on their parent 
clouds is, for now, negligible for all the clump sample. 
This is because these clumps are still in extremely early evolutionary 
stages of star formation and have not yet been severely affected  
by protostellar feedback (consistent with being 24 and 70 $\mu$m dark).   
Therefore, these clumps are ideal objects for studying the initial 
conditions of high-mass star and cluster formation.

\subsection{SiO Emission Unrelated to CO Outflows} 
\label{sec:SiO}
SiO has been widely used to trace shocked gas, such as those 
associated with protostellar outflows and jets 
\citep{1999ApJ...527L.117Z,2000AJ....119.1345Z,2006ApJ...636L.141H,
2006ApJ...636L.137P,2013A&A...550A..81C,2014A&A...570A..49L,
2019ApJ...878...29L,2019ApJ...886..130L}.
This is because the SiO formation is closely linked to the shock 
activities, which releases Si atoms from dust grains into gas phase 
through sputtering or vaporization \citep{1997A&A...321..293S}.

A total of 28 out of 43 CO outflows are detected in the SiO 5-4 
line (see  Figure~\ref{fig:outfmap}). 
The outflows seen in the SiO emission are more compact as compared 
with the CO spatial distribution, and appears to follow the outflow 
direction defined by CO.
In addition, the SiO emission is significantly weaker than 
the CO emission. This difference is most likely due to the 
higher excitation conditions (density, energy)   and different 
chemical conditions required by the SiO 5-4 transition with 
respect to the CO 2-1 transition 
\citep[e.g.,][]{2019ApJ...878...29L}. 
The comparison of outflow velocities measured by the CO line shows 
no difference between the outflows with SiO detection and without 
SiO detection. This indicates that the production of SiO requires 
other conditions (e.g., density and temperature) in addition to 
the appropriate velocity.

From Figure~\ref{fig:outfmap}, it is interesting to note that 
there is  strong SiO emission that is neither associated with CO 
outflows nor dense cores. As previously mentioned, SiO emission 
is closely related to the shock phenomena, therefore its presence 
suggests that there are strong shocks unrelated to star formation 
activities such as outflows or jets. Figure~\ref{fig:spectSiO} 
shows an overview of the SiO spectra toward G10.99. The SiO and 
CO emission, which is associated with protostellar outflows, 
appear to be comparable in the line central velocity 
(e.g., panel d, e, g and h of Figure~\ref{fig:spectSiO}). 
However, we note that there is also SiO emission that appears 
to be unrelated to outflows and dense cores.

Assuming optically thin and local thermodynamic equilibrium (LTE) 
condition for the SiO emission, the SiO column densities  
estimated to be 
$N$(SiO) =  $8.1\, \times \, 10^{11} - 5.9\, \times \, 10^{13}\,  \rm cm^{-2}$, 
with a median value of $4.7\, \times \, 10^{12}\,  \rm cm^{-2}$. 
The derived SiO column densities are comparable to that measured in 
high- and low-mass star forming regions \citep[e.g.,][]{1991A&A...243L..21B,
2006ApJ...636L.141H,2013ApJ...778...72F}.
To derive SiO abundances, we estimate the H$_{2}$ column densities 
from CO line (see Appendix \ref{app:equation}). 
The SiO abundance is found to be  
$X$(SiO) = $1.1\, \times \, 10^{-9} - 1.6\, \times \, 10^{-7}$ 
with a median value of $1.0\, \times \, 10^{-8}$.
% which is similar 
%to that measured in L1448mm, HH211 
%\citep{1991A&A...243L..21B,2006ApJ...636L.141H}, 
%and L1157 \citep{2000AJ....119.1345Z}.
Overall, we find no significant differences 
in the SiO column 
densities for the SiO gas components associated with and without 
CO outflows, except for G10.19, which shows relatively higher SiO 
column densities toward  the CO outflows as compared to the regions 
without CO outflow counterparts.

In most  cases the shocked regions traced by SiO emission without 
CO outflow counterpart are associated with multiple velocity 
components, as revealed by the CO emission (PV diagram and 
channel maps). 
In some cases, the velocities of the SiO emission are around 
the clump velocity  (see panel b and c of Figure~\ref{fig:spectSiO}),  
while some of them are offset from the clump velocity 
(see panel i of Figure~\ref{fig:spectSiO}). 
The interaction  between different velocity 
components can enhances the local shock activities, resulting an 
enhancement in the SiO emission \citep{2010MNRAS.406..187J}. 
Figure~\ref{fig:spectSiO} shows the PV diagram of the CO emission 
across the SiO contours for one sub-regions, the bright SiO emission 
appears to be at the location of a velocity shear (panel j in 
Figure~\ref{fig:spectSiO}); interfaces between different velocity 
components. The FWHM line widths of SiO measured toward these 
shocked regions are $\geqslant$ 5 \kms, which are  broader than the 
narrow line width of $\sim$0.8 \kms\ measured for the widespread SiO 
emission along the IRDC G035.39$-$0.33 \citep{2010MNRAS.406..187J}. 
The broad SiO emission is a further hint of gas movement due to the 
interaction between different gas components. This indicates  
that the shocks that produce this SiO emission most likely arise from 
 colliding or intersecting gas flows. For the other clumps, we also find 
that some of the SiO emission is located at the positions where 
velocity shears in the CO emission may be present. This SiO emission 
is most likely a result of collision or intersection of gas flows as well.

On the other hand, we find that some of the SiO emission regions show no 
signs of velocity shears in the CO emission (e.g., the regions corresponding 
to the spectra of panel b and c). 
In addition to collision/intersection of gas flows, the large/small-scale 
converging flows, the larger scale collapse, a population of undetected 
low-mass objects and some dense cores outside of the map could also 
create shocks, resulting in an enhancement 
in the SiO emission \citep[e.g.,][]{2010MNRAS.406..187J,2011ApJ...740L...5C,2013ApJ...773..123S,2014A&A...570A...1D}. 
Unfortunately, we can not distinguish between these possibilities with 
the data at hand. They can be tested by spatially resolving both large 
and small scale gas motions with reliable dense gas tracers 
(e.g, N$_{2}$H$^{+}$) or searching for deeply embedded distributed low-mass protostars in near-infrared observations as has been done in another IRDC by \cite{2014ApJ...791..108F}.

\section{Conclusions}
\label{sec:con}
In this paper, we analyzed the data from ASHES obtained with ALMA 
to study molecular outflow and shock  properties in 
12 massive 70 \um\ dark clumps.  
\begin{enumerate}
  \item  
  Outflow activities revealed by either CO or SiO emission are  
  found in 9 out of the 12 clumps observed. 
We successfully identified 43 molecular outflows in 41 out of 301  
dense cores, with an outflow detection rate of 14\% in the CO 2-1 emission.  
Among the 41 cores, 12 cores are associated with single bipolar outflows, 
2 cores host two bipolar outflows, while the remaining 27 cores have 
single uniploar outflows. 
The CO outflows are detected in both low-mass and high-mass cores. 
The maximum velocity of CO outflows reaches up to 95 \kms\ with respect 
to the systemic velocity.  
These results suggest that most of 
70~\um\ dark clumps already host  protostars, and thus they cannot be 
assumed to be prestellar without deep interferometric observations.

  \item 
The estimated protostellar accretion rates range from  2.0 $\times\, 10^{-9}$  
to 3.8 $\times\, 10^{-6}$ $M_{\odot}$~yr$^{-1}$ for the dense cores 
detected outflows, which is smaller than the typical value reported  
in more evolved cores.  
The comparison of the total accretion rates and 
$L_{\rm bol}/M_{\rm clump}$ 
in different evolutionary stages indicates that the accretion rates 
increase with the evolution of star formation. 
For this sample, the median ratio of outflow mass to core mass is about 
$8\,\times \,10^{-3}$, while the median ratio of total outflow mass to 
clump mass is about $2\, \times \,10^{-4}$. 

  \item 
We found a positive correlation between the outflow dynamical 
timescale and the gas mass of dense cores. The observed 
increase of outflow 
dynamical timescales towards the more massive cores indicates 
that the accretion history is longer in the more massive cores 
compared to the less massive cores.

  \item 
We identified six episodic molecular outflows associated to cores 
of all masses. 
This indicates that episodic outflows begin in the earliest stages 
of protostellar evolution for different masses of cores.
The computed dynamical timescales ($t_{\rm knot}$) of episodic 
outflow events are 
in the range $6 \times 10^{2}$ to  $2.5 \times 10^{4}$~yr with time 
span ($\triangle t_{\rm knot}$) between consecutive episodic outflow 
events of $3.6 \times 10^{2} - 7.2 \times 10^{3}$~yr. 
The $\triangle t_{\rm knot}$ is much longer than the value ($\sim$500~yr) 
found in more evolved cores in W43-MM1 \citep{2020A&A...636A..38N}, 
which indicates that the ejection episodicity timescale is likely 
not constant over time and accretion bursts may be more frequently 
at later stages of evolution in a cluster environment. 

  \item  
In ASHES, the mechanical energy of outflows is much smaller 
than values from 
more evolved high-mass star-forming regions, for which outflow 
mechanical energy is comparable to the kinetic energy arising from 
the internal gas motions. The total outflow luminosity is smaller than 
the turbulent dissipation rate in the majority of the clumps. 
This suggests that the outflow feedback can not replenish the 
internal turbulence in the majority of the sources at the current epoch. 
In addition, the disruptive impact of the outflows on their 
parent clouds is negligible for this sample.

  \item  
We found strong SiO emission not associated with protostellar 
outflows.  This finding suggests that protostellar outflows 
are not the sole mechanism responsible for SiO production and 
emission. Alternatively, other processes, such as large scale 
collisions or intersection of  converging flows, can give 
rise to shocks.

\end{enumerate}

\acknowledgments
This work is supported by the National Key R\&D Program of China 
(No. 2017YFA0402604) and the Natural Science Foundation of China 
under grants U1731237 and 11590783.
SL acknowledges support from the CfA pre-doctoral fellowship 
and Chinese Scholarship Council. P.S. was partially supported by 
a Grant-in-Aid for Scientific Research (KAKENHI Number 18H01259) 
of Japan Society for the Promotion of Science (JSPS). P.S. 
gratefully acknowledge the support
from the NAOJ Visiting Fellow Program to visit the National
Astronomical Observatory of Japan in November-December 2016. 
Data analysis was in part carried out on the open use data 
analysis computer system at the Astronomy Data Center (ADC) 
of the National Astronomical Observatory of Japan. This paper 
makes use of the following ALMA data: ADS/JAO. 
ALMA\#2015.1.01539.S. 
ALMA is a partnership of ESO (representing its member states), 
NSF (USA) and NINS (Japan), together with NRC (Canada), MOST 
and ASIAA (Taiwan), and KASI (Republic of Korea), in cooperation 
with the Republic of Chile. The Joint ALMA Observatory is operated 
by ESO, AUI/NRAO and NAOJ.  Data analysis was in part carried out 
on the open use data analysis computer system at the Astronomy 
Data Center (ADC) of the National Astronomical Observatory of Japan.

%\vspace{5mm}
\facilities{ALMA.}

 \software{
 CASA \citep{2007ASPC..376..127M}, 
 APLpy \citep{2012ascl.soft08017R}, 
 Astropy \citep{2013A&A...558A..33A},
 DS9 \citep{2003ASPC..295..489J},
 Matplotlib \citep{4160265}.
 }


\bibliography{IRDCoutflow}{}
\bibliographystyle{aasjournal}

%%% the collaboration and author truncation commands are used.  It has to


%\clearpage
\appendix
\section{Estimate of outflow parameters}
\label{app:equation}
Assuming that local thermodynamic equilibrium (LTE), we can estimate the 
CO column density ($N_{\rm CO}$),  outflow mass ($M_{\rm out}$), 
momentum ($P_{\rm out}$), energy ($E_{\rm out}$), based on the CO 
emission \citep{1983ApJ...265..824B,1992A&A...261..274C,
2015PASP..127..266M}: 
\begin{equation}
\label{outflow-mass}
N_{\rm CO} (\rm cm^{-2}) =  1.08 \times 10^{13} (T_{\rm ex} + 0.92)\; 
{\rm exp}\left(\frac{16.59}{T_{\rm ex}} \right)
\int \frac{\tau_{21}}{1-e^{\tau_{21}}} T_{\rm B}dv,
\end{equation}
\begin{equation}
\label{Mout}
M_{\rm out} = {d^{2}} \left[\frac{\rm H_{2}}{\rm CO}\right] 
\overline{m}_{\rm H_{2}} \int_{\Omega} N_{\rm CO}(\Omega) d\Omega,
\end{equation}
\begin{equation}
\label{Pout}
P_{\rm out} = M_{\rm r} v_{r} + M_{b} v_{\rm b},
\end{equation}
\begin{equation}
\label{Eout}
E_{\rm out} = \frac{1}{2} M_{\rm r} v_{\rm r}^{2} + \frac{1}{2} M_{\rm b} v_{\rm b}^{2},
\end{equation}
Using the CO outflow projected distance ($\lambda_{\rm max}$), 
we compute the outflow 
dynamical timescale ($t_{\rm dyn}$), outflow rate ($\dot{M}_{\rm out}$), 
outflow luminosity ($L_{\rm out}$), and mechanical force ($F_{\rm out}$): 
\begin{equation}
\label{tdyn}
t_{\rm dyn} = \frac{\lambda_{\rm max}}{(v_{\rm max(b)} + v_{\rm max(r)})/2},
\end{equation}
\begin{equation}
\label{Moutrate}
\dot{M}_{\rm out} = \frac{M_{\rm out}}{t_{\rm dyn}},
\end{equation} 
\begin{equation}
\label{Lout}
L_{\rm out} = \frac{E_{\rm out}}{t_{\rm dyn}},
\end{equation} 
\begin{equation}
\label{Fout}
F_{\rm out} = \frac{P_{\rm out}}{t_{\rm dyn}}.
\end{equation} 
Here,  $dv$ is the velocity interval in km~s$^{-1}$, $T_{\rm ex}$ is 
the line excitation temperature, $\tau_{21}$ is the optical 
depth, $T_{\rm B}$ is brightness temperature in K, $\Omega$ is the total 
solid angle that the flow subtends, $d$ is the source distance, 
$v_{\rm max(b)}$ and $v_{\rm max(r)}$ are  the maximum velocities of 
CO blue-shifted and red-shifted emission, respectively. $M_{\rm r}$ and 
$M_{\rm b}$ are the gas masse of CO outflows at the corresponding 
blue-shifted ($v_{\rm b}$) and red-shifted 
($v_{\rm r}$) velocities, respectively.

A range of excitation temperatures, from 10 to 60 K, has been used to 
calculate the column density in order to understand the effect of 
excitation temperatures on the CO column density.  The estimated CO 
column density agree within a factor of 1.5 in this temperature range, 
which indicates that the CO column density dose not significantly 
depend on the temperature. 
In this work, we  assume that  CO 
emission is optically thin in the line wing and that the dust 
temperature approximate the excitation temperature of outflow 
gas (see Table \ref{tab:clumps}), and adopt 
the CO-to-H$_{2}$ abundance of 10$^{-4}$, 
$\left[\frac{\rm H_{\rm 2}}{\rm C\rm O}\right]$ = 10$^{4}$ 
\citep{1987ApJ...315..621B}, and mean molecular mass per hydrogen 
molecule $\overline{m}_{\rm H_{2}}$ = 2.72m$_{\rm H}$. 
The  CO velocity integrated intensity has been corrected for primary 
beam attenuation, while the  inclination of the outflow 
axis with respect to the line of sight have not been 
corrected in the estimation of the outflow parameters.

%--------------------------------------------------------------------------------------
\floattable
\begin{deluxetable*}{cccccccccccccccc}
\tabletypesize{\scriptsize}
\rotate
\tablecolumns{13}
\tablewidth{0pc}
\tablecaption{Physical Properties of the Clumps and Observational Parameters. \label{tab:clumps}}
\tablehead{
\colhead{Name} &\colhead{abbreviation}  &\colhead{R.A.}   &\colhead{Decl.}   &\colhead{Dist.} 
&\colhead{$M_{\rm cl}$} &\colhead{$R_{\rm cl}$} &\colhead{$L_{\rm bol}$}	 &\colhead{$v_{\rm lsr}$}  
&\colhead{$\sigma$} 
&\colhead{$T_{dust}$} &\colhead{$\sum_{\rm cl}$}&\colhead{$n_{\rm cl}$} &\colhead{$\theta_{\rm cont}$} 
&\colhead{$\theta_{\rm CO}$} &\colhead{$\theta_{\rm SiO}$} \\
  &     &\colhead{(J2000)}   &\colhead{(J2000)}   &\colhead{(kpc)} 
&\colhead{($M_{\odot}$)} &\colhead{(pc [\arcsec])} &\colhead{(L$_{\odot}$)}  &\colhead{(km s$^{-1}$)}  
&\colhead{(km s$^{-1}$)} 
&\colhead{(K)}   &\colhead{(g cm$^{-2}$)} &\colhead{($10^{4}$ cm$^{-3}$)}  &\colhead{(\arcsec $\times$ \arcsec)}
&\colhead{(\arcsec $\times$ \arcsec)}&\colhead{(\arcsec $\times$ \arcsec)}
}
\decimalcolnumbers
\startdata
G010.991$-$00.082  & G10.99 & 18:10:06.65  & $-$19.27.50.7  & 3.7  & 2230 & 0.49 (27)  & 467 & 29.5  & 1.27  & 12.0 & 0.50 & 5.3 & 1.29 $\times$ 0.86  & 1.44	 $\times$	0.99 & 1.57	 $\times$	1.08	\\
G014.492$-$00.139   & G14.49 & 18:17:22.03  & $-$16.25.01.9  & 3.9  & 5200 & 0.44 (23)  & 1211 & 41.1  & 1.68  & 13.0 & 1.10 & 13.0 & 1.29 $\times$ 0.85   & 1.43	 $\times$	0.99 & 1.51	 $\times$	1.05	\\
G028.273$-$00.167  & G28.23 & 18:43:31.00  & $-$04.13.18.1  & 5.1  & 1520 & 0.59 (24)  & 402 & 80.0  & 0.81  & 12.0 & 0.28 & 2.4 & 1.28 $\times$ 1.20   & 1.44	 $\times$	1.36 & 1.50	 $\times$	1.43	\\
G327.116$-$00.294  & G327.11 & 15:50:57.18  & $-$54.30.33.6  & 3.9  & 580 & 0.39 (20)  & 408 & $-$58.9  & 0.56  & 14.3 & 0.26 & 3.5 & 1.32 $\times$ 1.11   & 1.51	 $\times$	1.26 & 1.60	 $\times$	1.34	\\
G331.372$-$00.116  & G331.27 & 16:11:34.10  & $-$51.35.00.1  & 5.4  & 1640  & 0.63 (24)  & 776 & $-$87.8  & 1.29  & 14.0 & 0.20 & 1.7 & 1.34 $\times$ 1.09   & 1.54	 $\times$	1.24 & 1.63	 $\times$	1.32	\\
G332.969$-$00.029  & G332.96 & 16:18:31.61  & $-$50.25.03.1  & 4.4  & 730  & 0.59 (28)  & 195 & $-$66.6  & 1.41  & 12.6 & 0.10 & 0.9 & 1.35 $\times$ 1.08   & 1.55	 $\times$	1.22 & 1.64	 $\times$	1.30	\\
G337.541$-$00.082  & G337.54 & 16:37:58.48  & $-$47.09.05.1  & 4.0  & 1180  & 0.42 (22)  & 293 & $-$54.6  & 2.01  & 12.0 & 0.40 & 5.0 & 1.29 $\times$ 1.18   & 1.45	 $\times$	1.34 & 1.56	 $\times$	1.45	\\
G340.179$-$00.242  & G240.17 & 16:48:40.88  & $-$45.16.01.1  & 4.1  & 1470  & 0.74 (37)  & 645 & $-$53.7  & 1.48  & 14.0 & 0.12 & 0.9 & 1.41 $\times$ 1.29   & 1.51	 $\times$	1.48 & 1.56	 $\times$	1.53	\\
G340.222$-$00.167  & G340.22 & 16:48:30.83  & $-$45.11.05.8  & 4.0  & 760  & 0.36 (19)  & 728 & $-$51.3  & 3.04  & 15.0 & 0.38 & 5.5 & 1.40 $\times$ 1.28   & 1.49	 $\times$	1.47 & 1.53	 $\times$	1.51	\\
G340.232$-$00.146  & G340.23 & 16:48:27.56  & $-$45.09.51.9  & 3.9  & 710  & 0.48 (25)  & 332 & $-$50.8  & 1.23  & 14.0 & 0.15 & 1.7 & 1.39 $\times$ 1.26   & 1.49	 $\times$	1.47 & 1.53	 $\times$	1.50	\\
G341.039$-$00.114  & G341.03 & 16:51:14.11  & $-$44.31.27.2  & 3.6  & 1070  & 0.47 (27)  & 634 & $-$43.0  & 0.97  & 14.3 & 0.26 & 2.9 & 1.30 $\times$ 1.18   & 1.47	 $\times$	1.34 & 1.70	 $\times$	1.54	\\
G343.489$-$00.416 & G343.48 & 17:01:01.19 & $-$42.48.11.0 & 2.9 & 810 & 0.42 (29) & 85 & $-$29.0 & 1.00 & 10.3 & 0.30 & 3.8 & 1.30 $\times$ 1.18  & 1.47	 $\times$	1.34 & 1.70	 $\times$	1.53	\\
\enddata
\tablenotetext{}{Notes. 
Columns (5) to (13) are the physical properties of clumps  \citep{2019ApJ...886..102S}. 
Columns (14) to (16) are the continuum image, CO 2-1 and SiO 5-4 synthesized beam size in the 
combined 12m, 7m and  TP data sets.
}
\end{deluxetable*}
%--------------------------------------------------------------------------------------

\begin{longrotatetable}
\begin{deluxetable*}{cccccccccccccc}
\tablecaption{Parameters of Identified CO Outflows.
\label{tab:outflow}}
\tablehead{
\colhead{Source}   &\colhead{Core ID}    &      &\colhead{$v_{\rm lsr}$} &\colhead{$\Delta v$}
&\colhead{$M_{\rm out}$} &\colhead{$P_{\rm out}$}  &\colhead{$E_{\rm out}$}  &\colhead{$\lambda_{\rm out}$} 
&\colhead{$t_{\rm dyn}$} &\colhead{$\dot{M}_{\rm out}$} &\colhead{$F_{\rm out}$}  &\colhead{$L_{\rm out}$} 
&\colhead{Confidence} 
\\
     &      &          &\colhead{(km s$^{-1}$)} &\colhead{(km s$^{-1}$)}
&\colhead{(10$^{-2}$ M$_{\odot}$)}&\colhead{(M$_{\odot}$ km s$^{-1}$)} &\colhead{(10$^{43}$ erg)}&\colhead{(pc)}
&\colhead{(10$^{4}$ yr)}   &\colhead{(10$^{-6}$ M$_{\odot}$ yr$^{-1}$)} &\colhead{(10$^{-4}$ M$_{\odot}$ km s$^{-1}$ yr$^{-1}$)}  
&\colhead{(10$^{32}$ erg s$^{-1}$)} 
}
%\cline{2-14}
\decimalcolnumbers
\startdata
\multirow{5 }{*}{G10.99} & \multirow{2 }{*}{1} & blue & 29.5 & [-29.9, 15.5] & 1.59 & 0.348 & 8.84 & 0.112 & 0.18 & 8.61 & 1.88 & 15.16 & M \\ 
				 & 				 & red & 29.5 & [42.6, 102.8] & 2.57 & 0.965 & 43.50 & 0.159 & 0.21 & 12.13 & 4.55 & 65.06 & M \\  \cline{2-14}
				 & 		{2} & blue & 30.1 & [17.2, 28.0] & 1.25 & 0.066 & 0.44 & 0.104 & 0.81 & 1.54 & 0.08 & 0.17 & M \\  \cline{2-14}
				 & 		{6} & blue & 26.9 & [1.9, 17.2] & 1.54 & 0.231 & 3.61 & 0.085 & 0.34 & 4.59 & 0.69 & 3.41 & D \\  \cline{2-14}
				 & 		{E2} & blue & 27.3 & [9.6, 16.5] & 1.11 & 0.160 & 2.35 & 0.075 & 0.41 & 2.69 & 0.39 & 1.80 & D \\ \hline
\multirow{10 }{*}{G14.49} & \multirow{2 }{*}{1} & blue & 38.6 & [9.8, 34.5] & 2.52 & 0.296 & 4.02 & 0.145 & 0.49 & 5.13 & 0.60 & 2.60 & D \\ 
				 & 				 & red & 38.6 & [42.1, 66.3] & 32.22 & 2.523 & 27.11 & 0.425 & 1.50 & 21.43 & 1.68 & 5.72 & D \\  \cline{2-14}
				 & \multirow{2 }{*}{2} & blue & 41.8 & [-27.0, 33.2] & 5.31 & 1.025 & 26.10 & 0.161 & 0.23 & 23.18 & 4.48 & 36.16 & D \\ 
				 & 				 & red & 41.8 & [47.8, 92.3] & 16.95 & 3.371 & 88.26 & 0.256 & 0.50 & 34.18 & 6.80 & 56.43 & D \\  \cline{2-14}
				 & \multirow{2 }{*}{3} & blue & 41.2 & [8.5, 33.9] & 9.52 & 1.127 & 14.11 & 0.169 & 0.51 & 18.77 & 2.22 & 8.83 & D \\ 
				 & 				 & red & 41.2 & [42.8, 62.5] & 2.16 & 0.138 & 1.48 & 0.097 & 0.45 & 4.84 & 0.31 & 1.05 & D \\  \cline{2-14}
				 & \multirow{2 }{*}{4} & blue & 40.3 & [-7.4, 34.5] & 10.95 & 1.585 & 32.15 & 0.127 & 0.26 & 41.95 & 6.07 & 39.06 & D \\ 
				 & 				 & red & 40.3 & [42.1, 97.5] & 29.72 & 1.820 & 10.05 & 0.346 & 0.59 & 50.29 & 3.08 & 5.39 & D \\  \cline{2-14}
				 & 		{14} & red & 40.2 & [42.8, 51.0] & 0.75 & 0.033 & 0.18 & 0.052 & 0.47 & 1.58 & 0.07 & 0.12 & D \\  \cline{2-14}
				 & 		{29} & blue & 39.1 & [20.6, 34.5] & 5.28 & 0.486 & 5.13 & 0.078 & 0.41 & 12.81 & 1.18 & 3.95 & D \\ \hline
\multirow{4 }{*}{G28.23} & \multirow{2 }{*}{1} & blue & 78.8 & [61.3, 74.6] & 3.05 & 0.227 & 1.95 & 0.204 & 1.14 & 2.68 & 0.20 & 0.54 & D \\ 
				 & 				 & red & 78.8 & [81.0, 88.6] & 4.02 & 0.160 & 0.74 & 0.191 & 1.90 & 2.12 & 0.08 & 0.12 & D \\  \cline{2-14}
				 & 		{6} & red & 80.0 & [83.5, 89.2] & 0.74 & 0.036 & 0.19 & 0.122 & 1.30 & 0.57 & 0.03 & 0.05 & D \\ \hline
\multirow{7 }{*}{G327.11} & \multirow{2 }{*}{1} & blue & -59.1 & [-79.4, -62.8] & 4.69 & 0.393 & 3.48 & 0.138 & 0.67 & 7.01 & 0.59 & 1.65 & D \\ 
				 & 				 & red & -59.1 & [-52.0, -41.9] & 0.22 & 0.030 & 0.41 & 0.053 & 0.30 & 0.74 & 0.10 & 0.43 & D \\  \cline{2-14}
				 & \multirow{3 }{*}{2} & blue1 & -59.3 & [-146.7, -62.8] & 1.53 & 0.030 & 2.14 & 0.091 & 0.10 & 14.96 & 0.30 & 6.65 & M \\ 
				 & 				 & red1 & -59.3 & [-50.8, -43.2] & 2.05 & 0.233 & 2.74 & 0.175 & 1.06 & 1.92 & 0.22 & 0.82 & M \\ 
				 & 				 & blue2 & -59.3 & [-132.1, -62.8] & 1.74 & 0.213 & 10.83 & 0.154 & 0.21 & 8.36 & 1.03 & 16.54 & M \\ 
				 & 				 & red2 & -59.3 & [-50.8, -12.7] & 3.82 & 0.791 & 18.77 & 0.223 & 0.47 & 8.17 & 1.69 & 12.74 & M \\  \cline{2-14}
				 & 		{3} & red & -59.6 & [-52.1, -43.2] & 0.18 & 0.022 & 0.28 & 0.043 & 0.26 & 0.71 & 0.09 & 0.35 & D \\  \cline{2-14}
				 & 		{4} & red & -59.6 & [-58.4, -51.4] & 0.95 & 0.034 & 0.15 & 0.037 & 0.45 & 2.12 & 0.08 & 0.10 & D \\ \hline
\multirow{9 }{*}{G331.37} & \multirow{2 }{*}{3} & blue & -88.0 & [-140.0, -92.4] & 14.77 & 1.910 & 35.89 & 0.210 & 0.40 & 37.35 & 4.83 & 28.77 & D \\ 
				 & 				 & red & -88.0 & [-82.9, -67.6] & 9.44 & 0.876 & 9.35 & 0.250 & 1.20 & 7.88 & 0.73 & 2.47 & D \\  \cline{2-14}
				 & 		{4} & blue & -88.0 & [-103.2, -94.3] & 5.44 & 0.567 & 6.22 & 0.159 & 1.03 & 5.30 & 0.55 & 1.92 & D \\  \cline{2-14}
				 & 		{5} & red & -88.0 & [-77.8, -70.8] & 0.56 & 0.068 & 0.85 & 0.113 & 0.64 & 0.87 & 0.11 & 0.42 & D \\  \cline{2-14}
				 & \multirow{2 }{*}{14} & blue & -87.5 & [-102.5, -93.0] & 6.09 & 0.609 & 6.51 & 0.268 & 1.74 & 3.49 & 0.35 & 1.18 & D \\ 
				 & 				 & red & -87.5 & [-77.8, -70.8] & 0.80 & 0.096 & 1.18 & 0.273 & 1.60 & 0.50 & 0.06 & 0.23 & D \\  \cline{2-14}
				 & 		{19} & blue & -88.0 & [-108.2, -98.1] & 2.62 & 0.465 & 8.30 & 0.129 & 0.51 & 5.10 & 0.90 & 5.11 & D \\  \cline{2-14}
				 & 		{29} & red & -88.0 & [-77.8, -70.8] & 0.56 & 0.071 & 0.91 & 0.172 & 0.98 & 0.58 & 0.07 & 0.30 & D \\ \hline
\multirow{6 }{*}{G337.54} & \multirow{2 }{*}{1} & blue & -54.5 & [-76.3, -55.9] & 11.65 & 0.673 & 5.74 & 0.200 & 0.92 & 12.64 & 0.73 & 1.97 & D \\ 
				 & 				 & red & -54.5 & [-52.1, -40.7] & 18.22 & 0.972 & 6.38 & 0.449 & 3.07 & 5.94 & 0.32 & 0.66 & D \\  \cline{2-14}
				 & \multirow{2 }{*}{2} & blue & -54.8 & [-69.3, -56.6] & 5.71 & 0.304 & 2.07 & 0.377 & 2.59 & 2.21 & 0.12 & 0.25 & M \\ 
				 & 				 & red & -54.8 & [-50.9, -40.7] & 3.04 & 0.196 & 1.45 & 0.313 & 2.14 & 1.42 & 0.09 & 0.21 & M \\  \cline{2-14}
				 & 		{3} & red & -54.5 & [-53.0, -40.7] & 1.52 & 0.071 & 0.48 & 0.099 & 0.73 & 2.08 & 0.10 & 0.21 & D \\  \cline{2-14}
				 & 		{9} & red & -55.6 & [-53.4, -40.1] & 3.08 & 0.126 & 0.84 & 0.120 & 0.84 & 3.67 & 0.15 & 0.32 & D \\ \hline
\multirow{2 }{*}{G340.23} & 		{1} & blue & -50.5 & [-68.8, -52.9] & 0.93 & 0.046 & 0.28 & 0.111 & 0.59 & 1.57 & 0.08 & 0.15 & D \\  \cline{2-14}
				 & 		{2} & blue & -50.7 & [-70.1, -58.0] & 1.44 & 0.132 & 1.30 & 0.144 & 0.73 & 1.99 & 0.18 & 0.57 & D \\ \hline
\multirow{14 }{*}{G341.03} & \multirow{4 }{*}{1} & blue1 & -46.1 & [-67.5, -46.6] & 4.89 & 0.184 & 1.15 & 0.315 & 1.44 & 3.39 & 0.13 & 0.25 & D \\ 
				 & 				 & red1 & -46.1 & [-42.1, -37.0] & 0.76 & 0.035 & 0.17 & 0.296 & 3.18 & 0.24 & 0.01 & 0.02 & D \\ 
				 & 				 & blue2 & -46.1 & [-89.7, -46.6] & 3.80 & 0.517 & 9.02 & 0.368 & 1.36 & 2.79 & 0.38 & 2.10 & D \\ 
				 & 				 & red2 & -46.1 & [-42.1, -4.2] & 4.09 & 0.939 & 26.94 & 0.187 & 0.44 & 9.40 & 2.16 & 19.63 & D \\  \cline{2-14}
				 & 		{2} & blue & -43.0 & [-54.2, -44.0] & 0.36 & 0.002 & 0.04 & 0.066 & 0.58 & 0.63 & 0.00 & 0.02 & D \\  \cline{2-14}
				 & 		{3} & blue & -43.3 & [-70.1, -49.1] & 3.66 & 0.663 & 12.54 & 0.421 & 1.54 & 2.38 & 0.43 & 2.58 & M \\  \cline{2-14}
				 & 		{4} & blue & -43.7 & [-69.4, -45.3] & 1.11 & 0.058 & 0.56 & 0.136 & 0.52 & 2.14 & 0.11 & 0.34 & D \\  \cline{2-14}
				 & 		{5} & blue & -43.3 & [-52.3, -46.6] & 0.35 & 0.017 & 0.09 & 0.108 & 1.17 & 0.30 & 0.01 & 0.02 & M \\  \cline{2-14}
				 & 		{6} & blue & -42.2 & [-56.1, -47.8] & 2.96 & 0.253 & 2.27 & 0.267 & 1.89 & 1.57 & 0.13 & 0.38 & M \\  \cline{2-14}
				 & \multirow{2 }{*}{13} & blue & -43.6 & [-138.6, -46.6] & 2.87 & 1.764 & 117.08 & 0.176 & 0.18 & 15.90 & 9.76 & 205.41 & D \\ 
				 & 				 & red & -43.6 & [-39.0, -15.5] & 0.92 & 0.059 & 0.39 & 0.173 & 2.09 & 0.44 & 0.03 & 0.06 & D \\  \cline{2-14}
				 & 		{14} & blue & -42.3 & [-70.7, -56.1] & 0.83 & 0.159 & 3.16 & 0.144 & 0.50 & 1.67 & 0.32 & 2.01 & D \\  \cline{2-14}
				 & 		{17} & blue & -42.9 & [-70.1, -49.1] & 0.75 & 0.105 & 1.73 & 0.067 & 0.24 & 3.14 & 0.44 & 2.29 & D \\  \cline{2-14}
				 & 		{32} & blue & -43.0 & [-59.9, -45.9] & 0.27 & 0.016 & 0.12 & 0.067 & 0.39 & 0.70 & 0.04 & 0.10 & M \\ \hline
\multirow{3 }{*}{G343.48} & 		{1} & red & -28.7 & [-19.8, -9.6] & 0.09 & 0.009 & 0.09 & 0.050 & 0.25 & 0.37 & 0.04 & 0.12 & M \\  \cline{2-14}
				 & 		{2} & red & -28.7 & [-21.0, -9.6] & 0.16 & 0.009 & 0.06 & 0.040 & 0.27 & 0.59 & 0.03 & 0.07 & D \\  \cline{2-14}
				 & 		{3} & red & -28.4 & [-24.9, -13.4] & 0.22 & 0.015 & 0.11 & 0.054 & 0.27 & 0.80 & 0.05 & 0.13 & D \\ \hline
Mean				 & 				 &  &  & 				 & 4.41 & 0.480 & 9.70 & 0.172 & 0.87 & 7.32 & 1.05 & 9.58 &  \\ \hline
Median				 & 				 &  &  & 				 & 2.16 & 0.196 & 2.14 & 0.145 & 0.58 & 2.69 & 0.30 & 0.82 &  \\ \hline
Minimum				 & 				 &  &  & 				 & 0.09 & 0.002 & 0.04 & 0.037 & 0.10 & 0.24 & 0.003  & 0.02 &  \\ \hline
Maximum				 & 				 &  &  & 				 & 32.22 & 3.371 & 117.08 & 0.449 & 3.18 & 50.29 & 9.76 & 205.41 &  \\ 
\enddata
\tablenotetext{}{Notes. 
Column (4): the core systemic velocity.
Column (5): velocity ranges of blue- and red-shifted components. 
Column (6): outflow mass. Column (7): outflow momentum. 
Column (8): outflow energy. Column (9): outflow lobes physical length. 
Column (10): outflow dynamical time. Column (11): outflow mass-loss rate. 
Column (12): mechanical force. Column (13): outflow luminosity. 
(14): confidence rating of outflow, D and M represent ``Definitive" and ``Marginal", respectively. 
The last four rows are the mean, median, minimum and maximum values of estimated parameters. 
The outflow parameters are not corrected for the inclination angle of outflows. 
}
\end{deluxetable*}
\end{longrotatetable}

\floattable
\begin{deluxetable*}{ccccccccccccccc}
\tabletypesize{\scriptsize}
\rotate
\tablecolumns{15}
\tablewidth{0pc}
\tablecaption{Assessment of Outflow Impact on Clumps. 
\label{tab:impact}}
\tablehead{
\colhead{Name}      &\colhead{$E_{\rm out}$} &\colhead{$L_{\rm out}$}   &\colhead{$E_{\rm grav} $} &
\colhead{$E_{\rm turb} $}  &\colhead{$t_{\rm diss}$} &\colhead{$\dot{E}_{\rm turb}$} &\colhead{$r_{L}$}    &  &
\colhead{$t_{\rm diss}$}   &\colhead{$\dot{E}_{\rm turb}$} &\colhead{$r_{L}$} &\colhead{$E_{\rm out}/E_{\rm grav}$} &
\colhead{$E_{\rm out}/E_{\rm turb}$}	 &\colhead{$M_{\rm out}/M_{\rm clump}$} \\
         &\colhead{(10$^{44}$ erg)} &\colhead{(10$^{33}$ erg s$^{-1}$)}  &\colhead{(10$^{46}$ erg)} &
\colhead{(10$^{46}$ erg)}  &\colhead{(10$^{5}$ yr)}  &\colhead{(10$^{33}$ erg s$^{-1}$)}   &       & 
&\colhead{(10$^{5}$ yr)}   &\colhead{(10$^{33}$ erg s$^{-1}$)}  & &   & \\
\cline{6-8} \cline{10-12} 
\colhead{(1)}  &\colhead{(2)}     &\colhead{(3)}      &\colhead{(4)} &
\colhead{(5)}   &\colhead{(6)}    &\colhead{(7)}  &\colhead{(8)}     &
   &\colhead{(9)} &\colhead{(10)} &\colhead{(11)} &
\colhead{(12)} &\colhead{(13)} &\colhead{(14)}
}
\startdata
G10.99  & 5.87 & 4.69 & 86.82 & 10.72 & 1.89 & 18.00 & 0.261 & & 0.39 & 86.21 & 0.054 & 6.76E-04 & 5.48E-03 & 3.62E-05 \\
G14.49   & 20.86 & 8.88 & 525.74 & 43.76 & 1.28 & 108.26 & 0.082 & & 1.52 & 91.22 & 0.097 & 3.97E-04 & 4.77E-03 & 2.22E-04 \\
G28.23  & 0.29 & 0.03 & 33.50 & 2.97 & 3.57 & 2.64 & 0.013 & & 0.48 & 19.79 & 0.002 & 8.59E-05 & 9.70E-04 & 5.14E-05 \\
G327.11  & 3.88 & 2.03 & 7.38 & 0.54 & 3.42 & 0.50 & 4.053 & & 0.45 & 3.76 & 0.539 & 5.26E-03 & 7.19E-02 & 2.62E-04 \\
G331.37  & 6.92 & 2.00 & 36.52 & 8.13 & 2.39 & 10.79 & 0.186 & & 0.24 & 109.49 & 0.018 & 1.90E-03 & 8.51E-03 & 2.46E-04 \\
G332.96  & … & … & 7.73 & 4.33 & 2.05 & 6.70 & … & & … & … & … & … & … & … \\
G337.54  & 1.57 & 0.16 & 28.36 & 14.22 & 1.02 & 44.09 & 0.004 & & 0.49 & 92.91 & 0.002 & 5.54E-04 & 1.11E-03 & 2.92E-04 \\
G340.17  & … & … & 24.98 & 9.60 & 2.45 & 12.44 & … & & … & … & … & … & … & … \\
G340.22  & … & … & 13.73 & 20.95 & 0.58 & 114.66 & … & & … & … & … & … & … & … \\
G340.23  & 0.16 & 0.07 & 8.98 & 3.20 & 1.91 & 5.31 & 0.013 & & 0.13 & 79.74 & 0.001 & 1.76E-04 & 4.93E-04 & 3.35E-05 \\
G341.03  & 17.53 & 8.43 & 20.84 & 3.00 & 2.37 & 4.01 & 2.103 & & 0.51 & 18.50 & 0.455 & 8.41E-03 & 5.85E-02 & 2.58E-04 \\
G343.48 & 0.03 & 0.03 & 13.36 & 2.41 & 2.06 & 3.72 & 0.008 & & 0.14 & 53.83 & 0.001 & 1.99E-05 & 1.10E-04 & 5.84E-06 \\
\enddata
\tablenotetext{}{Notes. 
Column (2)-(3): the total outflow energy and total outflow luminosity 
estimated by the CO emission.   
Column (4): the gravitational binding energy of the clump. 
Column (5): the turbulent energy of the clump. 
Column (6)-(8): the turbulent dissipation rate ($\dot{E}_{\rm turb}$) and 
ratio of outflow luminosity to turbulent dissipation rate 
($r_{L} = L_{\rm out}/\dot{E}_{\rm turb}$) 
are calculated using the dissipation time, column(6), derived from 
the first approach (Equation~\ref{Tdis0}).
Column (9)-(11): the $\dot{E}_{\rm turb}$ and  $r_{L}$ are 
computed using the dissipation time, 
column(9), derived from the second approach (Equation~\ref{Tdis1}). 
Column (12): the ratio of total outflow energy to gravitational binding energy. 
Column (13): the ratio of total outflow energy to turbulent energy. 
Column (14): the ratio of total outflow mass to clump mass. 
}
\end{deluxetable*}
%\end{longrotatetable}

%--------------------------------------------------------------------------------------
%\floattable
\begin{deluxetable*}{ccccccc}
\tabletypesize{\scriptsize}
%\rotate
\tablecolumns{15}
\tablewidth{0pc}
\tablecaption{Parameters of CO Knots. \label{tab:knots}}
\tablehead{
\colhead{Clump}   &\colhead{Core ID}  &\colhead{Knot}  &\colhead{$\lambda_{\rm knot}$}   
&\colhead{$v_{\rm knot}$} &\colhead{$t_{\rm knot}$} &\colhead{$\triangle v_{\rm knot}$} \\
&   &&\colhead{(pc)}   
&\colhead{(km s$^{-1}$)} &\colhead{(10$^{3}$ yr)} &\colhead{(10$^{3}$ yr)} 
}
\decimalcolnumbers
\startdata
\multirow{ 8 }{*}{ G14.49 } & \multirow{ 6 }{*}{ 1 } & B0 & 0.11  & 11.50  & 9.75  & ... \\ 
     &      & R0 & 0.07  & 10.50  & 6.41  & ... \\ 
     &      & R1 & 0.14  & 11.00  & 12.23  & 5.83  \\ 
     &      & R2 & 0.21  & 10.80  & 19.38  & 7.15  \\ 
     &      & R3 & 0.28  & 11.00  & 24.81  & 5.42  \\ 
     &      & R4 & 0.39  & 15.00  & 25.17  & 0.36  \\  \cline{2-7}
     & \multirow{ 3 }{*}{ 2 } & B0 & 0.14  & 60.00  & 2.24  & ... \\ 
     &      & R0 & 0.21  & 52.50  & 3.84  & ... \\
     &      & R1 & 0.25  & 43.00  & 5.65  & 1.81 \\   \hline
\multirow{ 2 }{*}{ G28.23 } & \multirow{ 2 }{*}{ 1 } & B0 & 0.04  & 9.76  & 4.46  & ... \\ 
     &      & B1 & 0.15  & 14.26  & 10.01  & 5.55  \\ \hline
\multirow{ 3 }{*}{ G331.37 } & \multirow{ 3 }{*}{ 3 } & B0 & 0.13  & 37.93  & 3.39  & ... \\ 
     &      & R0 & 0.05  & 20.57  & 2.50  & ... \\ 
     &      & R1 & 0.07  & 18.07  & 3.70  & 1.20  \\ \hline
\multirow{ 2 }{*}{ G340.23 } & \multirow{ 2 }{*}{ 2 } & B0 & 0.03  & 13.31  & 1.97  & ... \\ 
     &      & B1 & 0.11  & 16.11  & 6.96  & 5.00  \\ \hline
\multirow{ 3 }{*}{ G341.03 } & \multirow{ 3 }{*}{ 1 } & B0 & 0.02  & 36.45  & 0.56  & ... \\ 
     &      & B1 & 0.17  & 24.95  & 6.83  & 6.26  \\ 
     &      & B2 & 0.35  & 28.45  & 12.09  & 5.27  \\ \hline
   Mean  &      &  & 0.15  & 23.43  & 8.52  & 4.38  \\ \hline
   Median  &      &  & 0.14  & 16.11  & 6.41  & 5.35  \\ \hline
   Minimum  &      &  & 0.02  & 9.76  & 0.56  & 0.36  \\ \hline
   Maximum  &      &  & 0.39  & 60.00  & 25.17  & 7.15  \\ 
\enddata
\tablenotetext{}{Notes. 
Column (3): the name of knots.   
Column (4): the projected distance of knots. 
Column (5): the knots maximum velocity  respect to the core velocity. 
Column (6): the knot dynamical timescale. 
Column (7): the time span between consecutive knots. 
The last four rows are the mean, median, minimum, and maximum values of 
estimated parameters. 
The parameters are not corrected for the inclination angle of outflows. 
}
\end{deluxetable*}
%--------------------------------------------------------------------------------------

%---------------------
%---------------------
\begin{figure*}%\ContinuedFloat % Example image
\center
\subfloat[]{
 \includegraphics[scale=0.5]{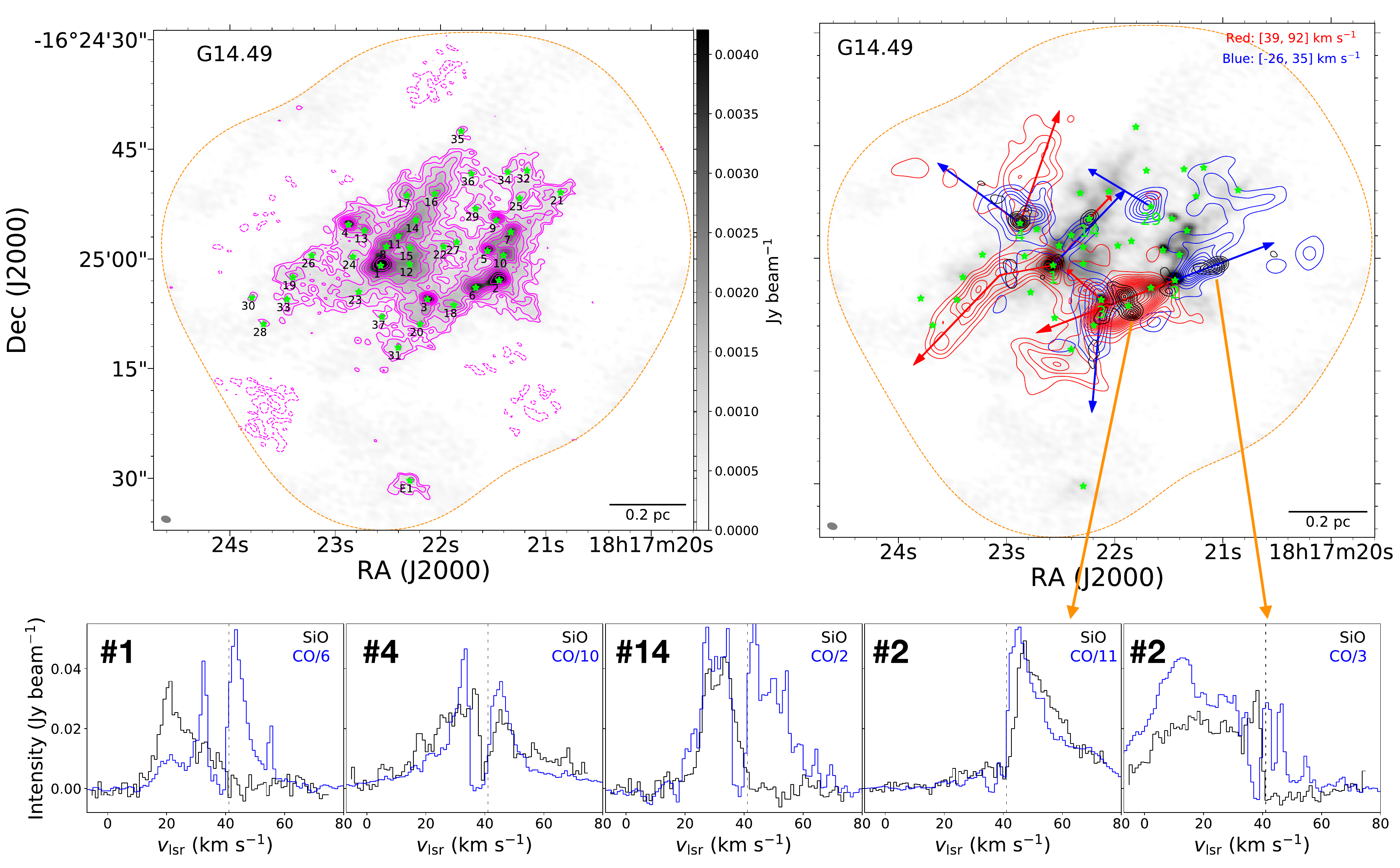}
}\\
\caption{Presented here are individual plots for Figure \ref{fig:outfmap} 
which are available as the online material. 
Dashed orange contour shows 20\% of the sensitivity level of the 
mosaic in the ALMA continuum image.  
The green stars indicate the identified dense cores. 
Upper left: ALMA 1.3~mm continuum image of G14.49. The contour 
levels are (-4, -3, 3, 4, 6, 8, 10, 12, 15,18, 25, 35)$\times \sigma$, with 
$\sigma = $0.168 mJy beam$^{-1}$.   
Upper right: The blue-shifted (blue contours) and red-shifted (red contours) 
components of CO emission, and SiO velocity-integrated intensity 
(black contours, integration range is between 3 and 75 \kms) 
 overlaid on the continuum emission.  The blue and 
red contours levels are (7, 12, 17, 22, 27 ....)$\times \sigma$, 
where the $\sigma$ are 0.1 Jy beam$^{-1}$ km s$^{-1}$ and 
0.12  Jy beam$^{-1}$ km s$^{-1}$ for blue and red contours, 
respectively.  The black contours are (3, 5, 7, 9, 11, 13, 15, 
17)$\times \sigma$, with $\sigma =$ 0.06 Jy beam$^{-1}$ km s$^{-1}$. 
Green stars show the position of the dust cores defined in 
\cite{2019ApJ...886..102S}. On the left side panels, black numbers 
correspond to the name of all cores. On the right side panels, the 
green numbers indicate only the cores associated with molecular outflows.
Bottom: example spectra of SiO and CO lines. The black vertical 
dashed lines represent the $v_{\rm lsr}$ of the clump.  
}
\label{fig:outfmap2}
\end{figure*}

\begin{figure*}\ContinuedFloat % Example image
\center
\subfloat[]{
 \includegraphics[scale=0.5]{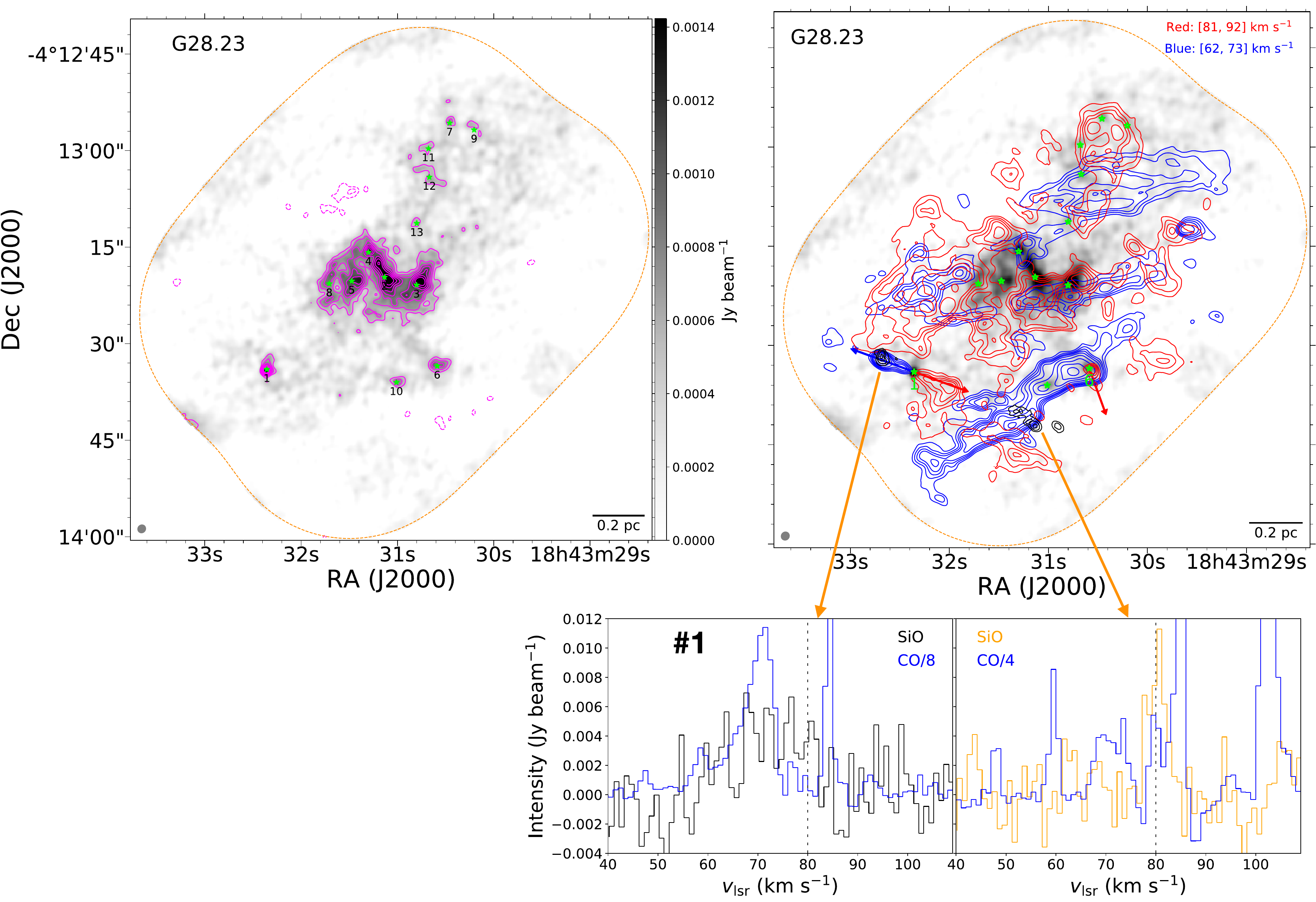}
}\\
\caption{Presented here are individual plots for Figure \ref{fig:outfmap} 
which are available as the online material.
Dashed orange contour indicates 20\% of the sensitivity level of the 
mosaic from the ALMA continuum image.
The green stars indicate the identified dense cores. 
Upper left: ALMA 1.3~mm continuum image of G28.23. The contour levels 
are (-4, -3, 3, 4, 5, 6, 7, 8, 9)$\times \sigma$, with 
$\sigma = $0.164 mJy beam$^{-1}$.  
Upper right: The blue-shifted (blue contours) and red-shifted (red contours) 
components of CO emission, and SiO velocity-integrated intensity 
(black contours, integration range is between 67 and 85 \kms)  overlaid on the 
continuum emission.  The blue and 
red contours levels are (4, 6, 9, 13, 18, 24, 31, 39, 49, 60)$\times \sigma$, 
where the $\sigma$ are 0.035 Jy beam$^{-1}$ km s$^{-1}$ and 
0.06  Jy beam$^{-1}$ km s$^{-1}$ for blue and red contours, respectively.  
The black contours are (3, 4, 5, 6, 7, 8, 9)$\times \sigma$,
with $\sigma =$ 0.015 Jy beam$^{-1}$ km s$^{-1}$. 
Green stars show the position of the dust cores. 
On the left side panels, black numbers 
correspond to the name of all cores. On the right side panels, the 
green numbers indicate only the cores associated with molecular outflows.
Bottom: example spectra of SiO and CO lines.  The black vertical 
dashed lines represent the $v_{\rm lsr}$ of the clump. 
}
%\label{fig:outfmap2}
\end{figure*}

\begin{figure*}\ContinuedFloat % Example image
\center
\subfloat[]{
 \includegraphics[scale=0.5]{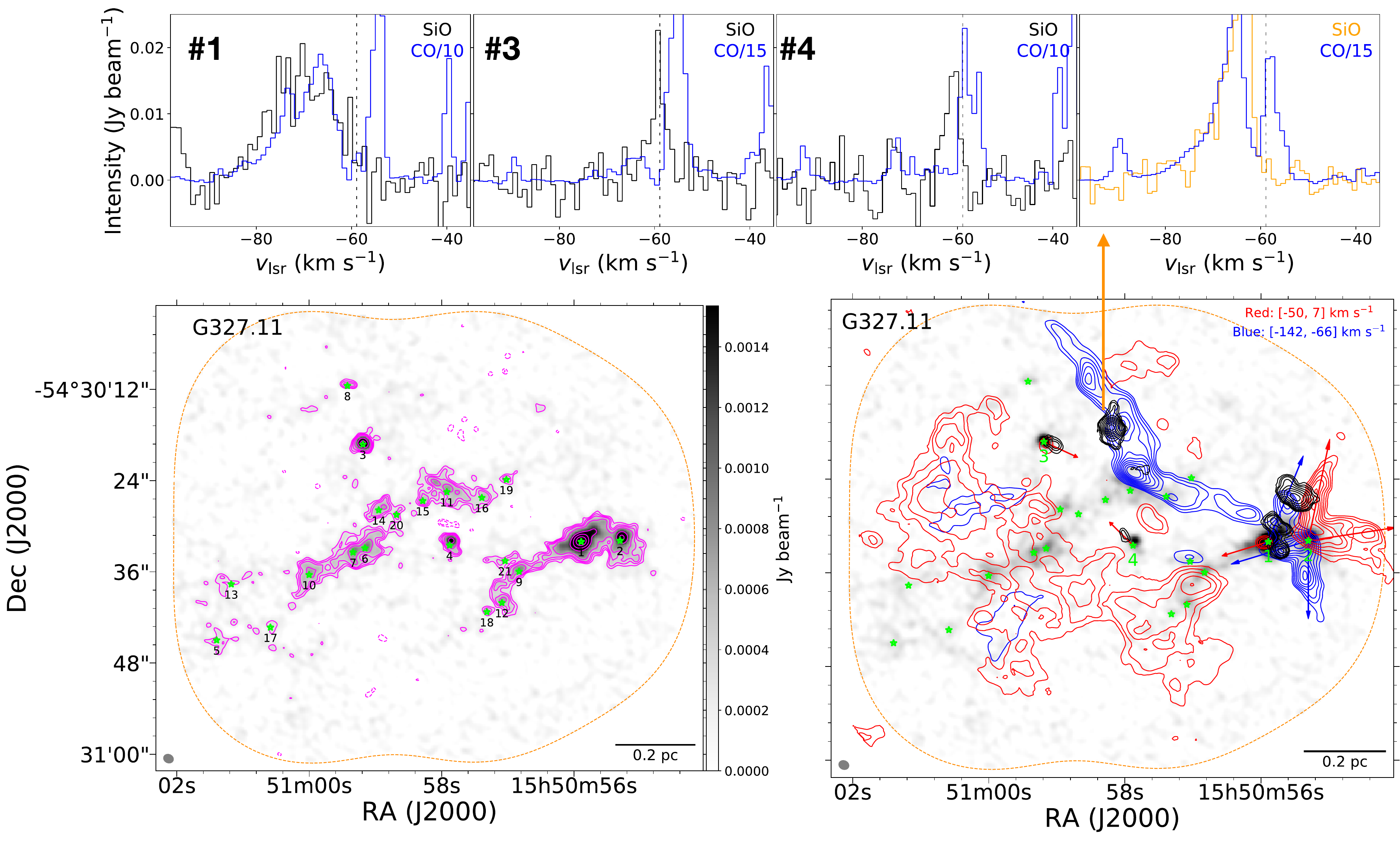}
}\\
\caption{Presented here are individual plots for Figure \ref{fig:outfmap} 
which are available as the online material. 
Dashed orange contour indicates 20\% of the sensitivity level of the 
mosaic from the ALMA continuum image.
The green stars indicate the identified dense cores. 
Upper: example spectra of SiO and CO lines.  The black vertical 
dashed lines represent the $v_{\rm lsr}$ of the clump. 
Bottom left: ALMA 1.3~mm continuum image of G327.11. The contour levels 
are (-4, -3, 3, 4, 5, 7, 10, 15, 23, 35)$\times \sigma$, with 
$\sigma = $0.089 mJy beam$^{-1}$.  
Bottom right: The blue-shifted (blue contours) and red-shifted (red contours) 
components of CO emission, and SiO velocity-integrated intensity 
(black contours, integration range is between -68 and -58 \kms)  
overlaid on the continuum emission.  The blue and 
red contours levels are (4, 6, 9, 13, 18, 24, 31, 39, 49, 60)$\times \sigma$, 
where the $\sigma$ are 0.04 Jy beam$^{-1}$ km s$^{-1}$ and 
0.025 Jy beam$^{-1}$ km s$^{-1}$ for blue and red contours, respectively.  
The black contours are (3, 4, 5, 7, 10, 14, 19, 25)$\times \sigma$,
with $\sigma =$ 0.01 Jy beam$^{-1}$ km s$^{-1}$. 
Green stars show the position of the dust cores. 
On the left side panels, black numbers 
correspond to the name of all cores. On the right side panels, the 
green numbers indicate only the cores associated with molecular outflows.
The directions of outflow \#1 and \#2 toward \#2--G327.11 are approximately 
along southeast-northwest and south-north, respectively. 
}
\end{figure*}

\begin{figure*}\ContinuedFloat % Example image
\center
\subfloat[]{
 \includegraphics[scale=0.5]{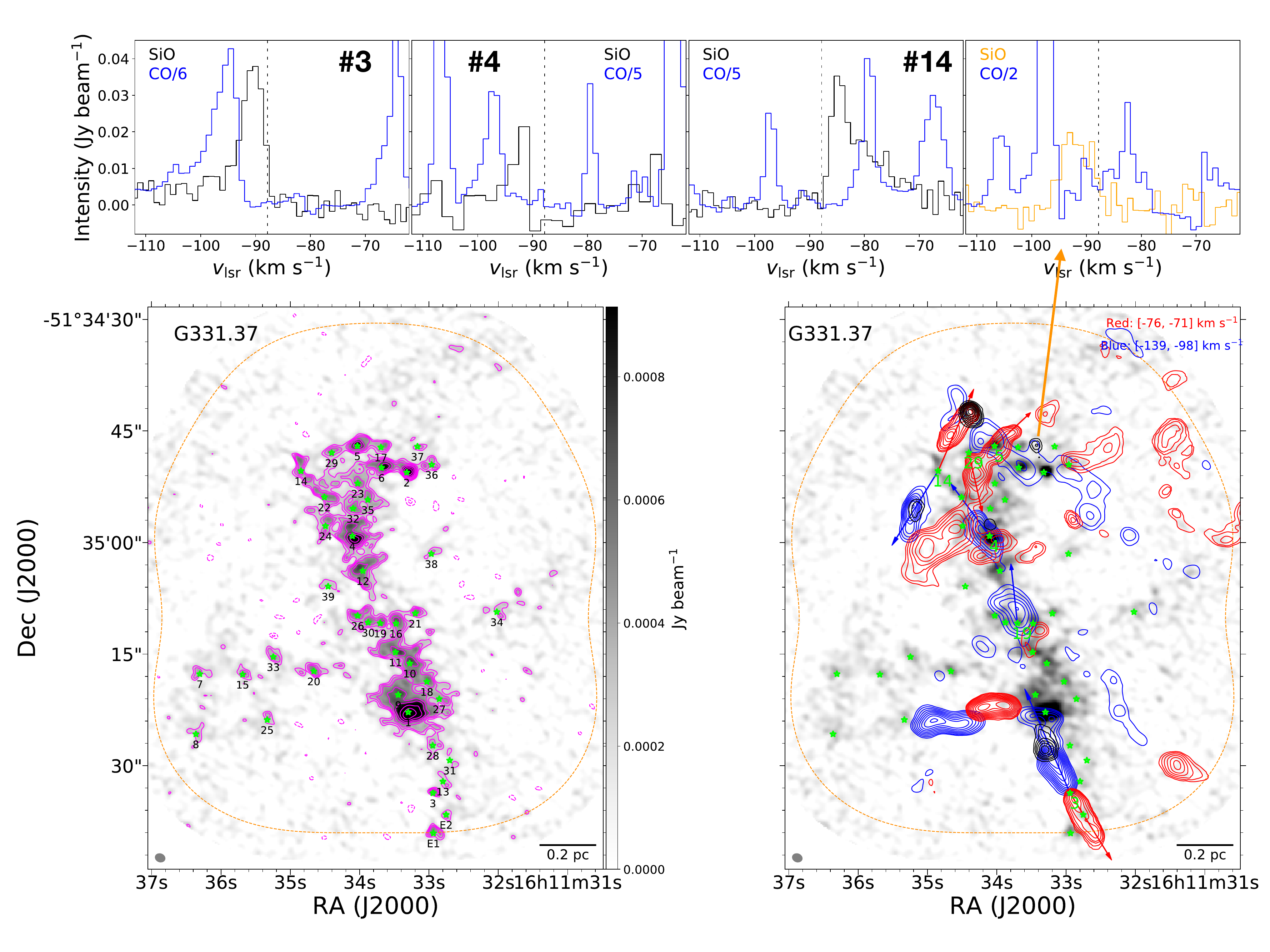}
}\\
\caption{Presented here are individual plots for Figure \ref{fig:outfmap} 
which are available as the online material.
Dashed orange contour shows 20\% of the sensitivity level of the 
mosaic from the ALMA continuum image.
The green stars indicate the identified dense cores. 
Upper: example spectra of SiO and CO lines.  The black vertical 
dashed lines represent the $v_{\rm lsr}$ of the clump. 
Bottom left: ALMA 1.3~mm continuum image of G331.37. The contour levels 
are (-4, -3, 3, 4, 5, 7, 9, 12, 16)$\times \sigma$, with 
$\sigma = $0.083 mJy beam$^{-1}$.  
Bottom right: The blue-shifted (blue contours) and red-shifted (red contours) 
components of CO emission, and SiO velocity-integrated intensity 
(black contours, integration range is between -95 and -69 \kms)  overlaid on the 
continuum emission.  The blue and 
red contours levels are (4, 6, 9, 13, 18, 24, 31, 39, 49, 60)$\times \sigma$, 
where the $\sigma$ are 0.025 Jy beam$^{-1}$ km s$^{-1}$ and 
0.01  Jy beam$^{-1}$ km s$^{-1}$ for blue and red contours, respectively.  
The black contours are (3, 4, 5, 6, 7, 10, 14, 19, 25)$\times \sigma$,
with $\sigma =$ 0.015 Jy beam$^{-1}$ km s$^{-1}$. 
Green stars show the position of the dust cores. 
On the left side panels, black numbers 
correspond to the name of all cores. On the right side panels, the 
green numbers indicate only the cores associated with molecular outflows.
}
\end{figure*}

\begin{figure*}\ContinuedFloat % Example image
\center
\subfloat[]{
 \includegraphics[scale=0.5]{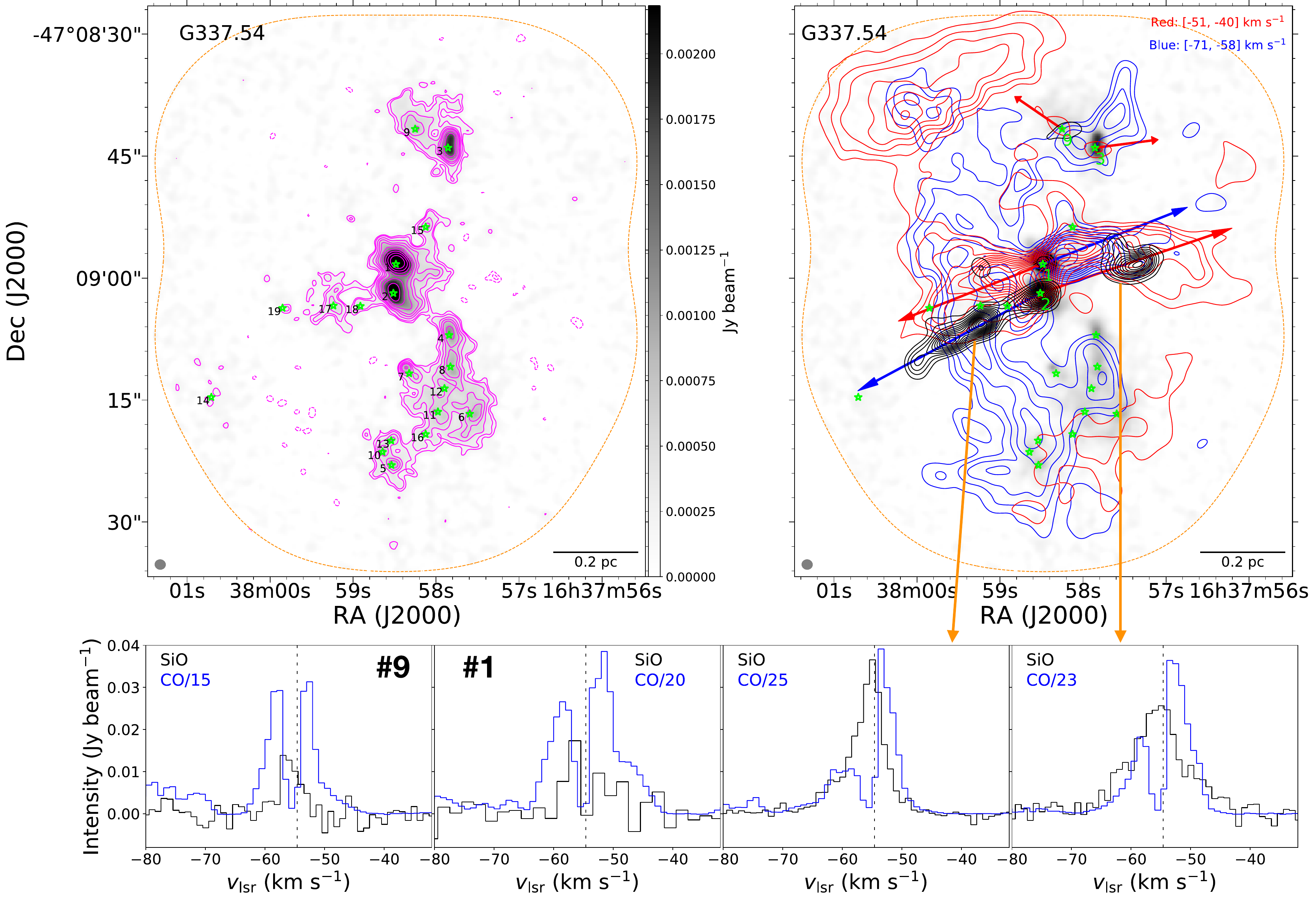}
}\\
\caption{Presented here are individual plots for Figure \ref{fig:outfmap} 
which are available as the online material.
Dashed orange contour shows 20\% of the sensitivity level of the 
mosaic from the ALMA continuum image.
The green stars indicate the identified dense cores. 
Upper left: ALMA 1.3~mm continuum image of G337.54. The contour levels 
are (-4, -3, 3, 4, 6, 8, 10, 14, 20, 30, 45, 75)$\times \sigma$, with 
$\sigma = $0.068 mJy beam$^{-1}$.  
Upper right: The blue-shifted (blue contours) and red-shifted (red contours) 
components of CO emission, and SiO velocity-integrated intensity 
(black contours, integration range is between -67 and -45 \kms)  overlaid on the 
continuum emission.  The blue and 
red contours levels are (5, 7, 9, 14, 20, 27, 35, 44, 54, 74)$\times \sigma$, 
where the $\sigma$ are 0.08 Jy beam$^{-1}$ km s$^{-1}$ and 
0.13  Jy beam$^{-1}$ km s$^{-1}$ for blue and red contours, respectively.  
The black contours are (3, 5, 6, 7, 9, 11, 13, 15, 17)$\times \sigma$,
with $\sigma =$ 0.015 Jy beam$^{-1}$ km s$^{-1}$. 
Green stars show the position of the dust cores. 
On the left side panels, black numbers 
correspond to the name of all cores. On the right side panels, the 
green numbers indicate only the cores associated with molecular outflows.
Bottom: example spectra of SiO and CO lines.  The black vertical 
dashed lines represent the $v_{\rm lsr}$ of the clump. 
}
\end{figure*}

\begin{figure*}\ContinuedFloat % Example image
\center
\subfloat[]{
 \includegraphics[scale=0.5]{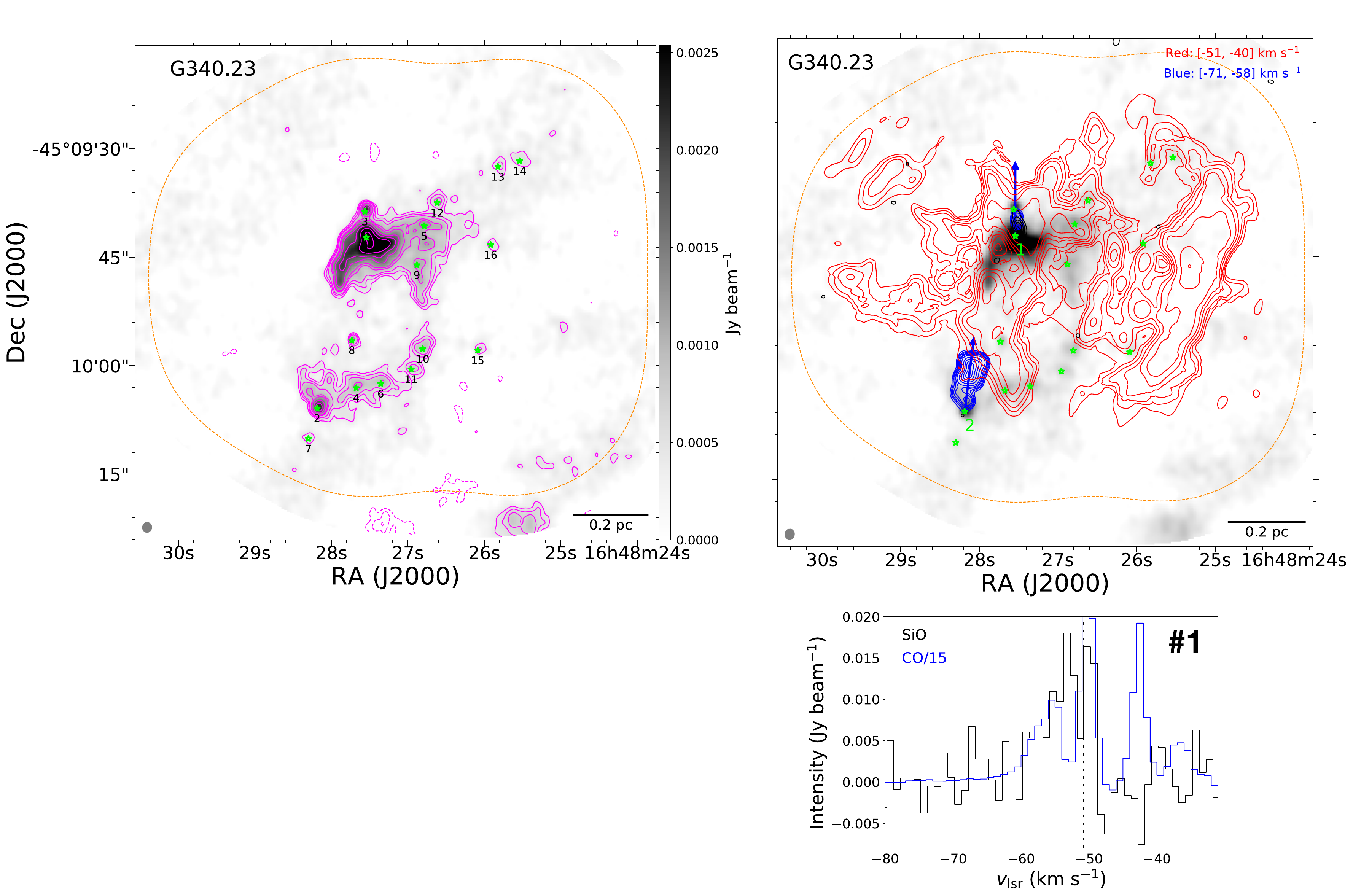}
}\\
\caption{Presented here are individual plots for Figure \ref{fig:outfmap} 
which are available as the online material.
Dashed orange contour indicates 20\% of the sensitivity level of the 
mosaic from the ALMA continuum image.
The green stars indicate the identified dense cores. 
Upper left: ALMA 1.3~mm continuum image of G340.23. The contour levels 
are (-4, -3, 3, 4, 5, 7, 8, 11, 14, 18, 23)$\times \sigma$, with 
$\sigma = $0.139 mJy beam$^{-1}$.  
Upper right: The blue-shifted (blue contours)  
components of CO emission and SiO velocity-integrated intensity 
(black contours, integration range is between -60 and -47 \kms)  overlaid on the 
continuum emission.  The blue  
red contours levels are (5, 7, 9, 12, 15, 20, 31, 30, 40, 50)$\times \sigma$, 
where the $\sigma$ are 0.02 Jy beam$^{-1}$ km s$^{-1}$ 
 for blue  contours.  
The black contours are (3, 4, 5, 6, 7, 10, 14, 19, 25)$\times \sigma$,
with $\sigma =$ 0.015 Jy beam$^{-1}$ km s$^{-1}$. 
Green stars show the position of the dust cores. 
On the left side panels, black numbers 
correspond to the name of all cores. On the right side panels, the 
green numbers indicate only the cores associated with molecular outflows.
Bottom: example spectra of SiO and CO lines.  The black vertical 
dashed lines represent the $v_{\rm lsr}$ of the clump. 
}
\end{figure*}

\begin{figure*}\ContinuedFloat % Example image
\center
\subfloat[]{
 \includegraphics[scale=0.5]{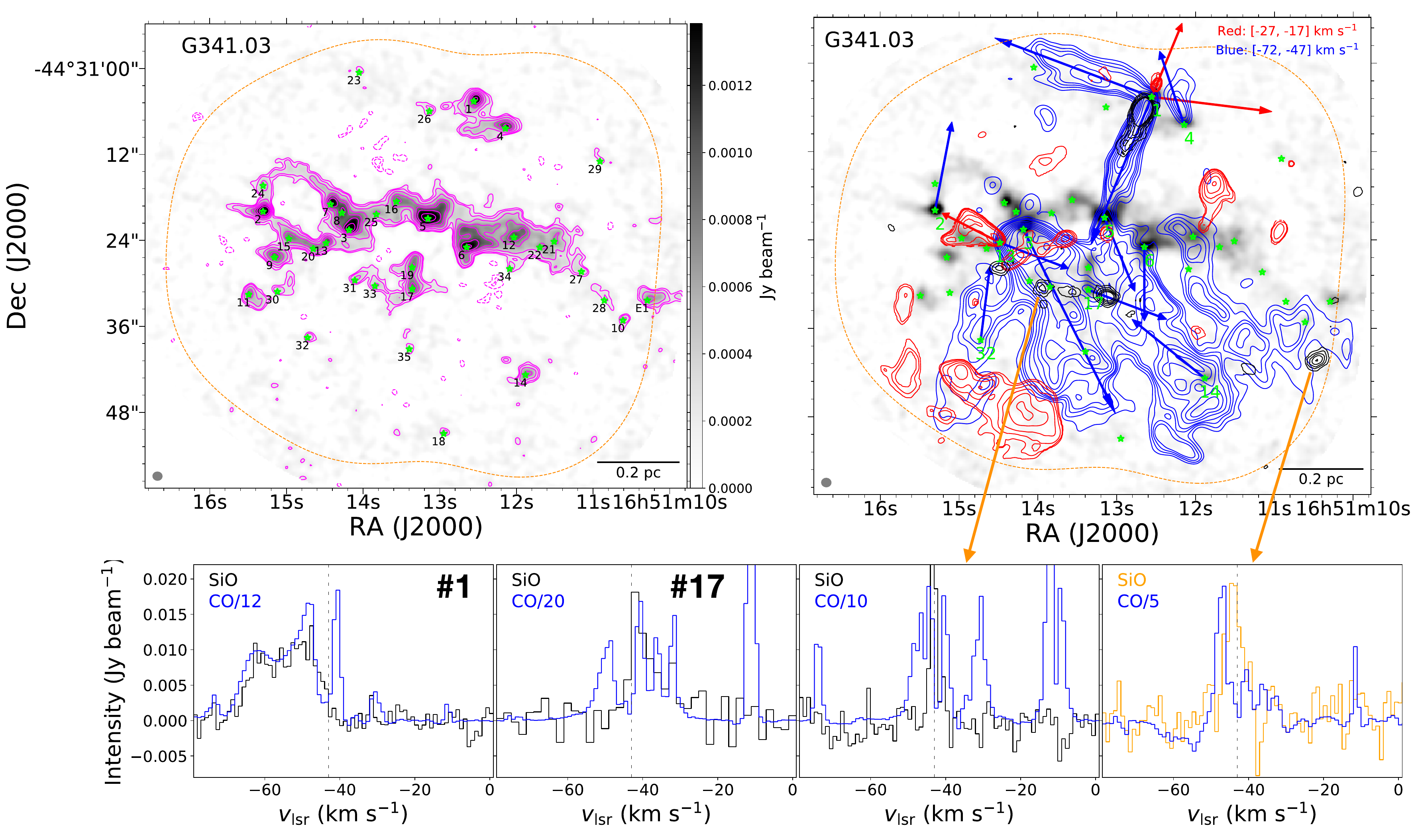}
}\\
\caption{Presented here are individual plots for Figure \ref{fig:outfmap} 
which are available as the online material.
Dashed orange contour indicates 20\% of the sensitivity level of the 
mosaic from the ALMA continuum image. 
The green stars indicate the identified dense cores. 
Upper left: ALMA 1.3~mm continuum image of G341.03. The contour levels 
are (-4, -3, 3, 4, 6, 9, 12, 16, 22)$\times \sigma$, with 
$\sigma = $0.070 mJy beam$^{-1}$.  
Upper right: The blue-shifted (blue contours) and red-shifted (red contours) 
components of CO emission, and SiO velocity-integrated intensity 
(black contours, integration range is between -66 and -31 \kms)  overlaid on the 
continuum emission.  The blue and 
red contours levels are (5, 7, 9, 14, 20, 27, 35, 44, 54, 74)$\times \sigma$, 
where the $\sigma$ are 0.07 Jy beam$^{-1}$ km s$^{-1}$ and 
0.015  Jy beam$^{-1}$ km s$^{-1}$ for blue and red contours, respectively.  
The black contours are (3, 5, 6, 7, 9, 11, 13, 15, 17)$\times \sigma$,
with $\sigma =$ 0.02 Jy beam$^{-1}$ km s$^{-1}$. 
Green stars show the position of the dust cores. 
On the left side panels, black numbers 
correspond to the name of all cores. On the right side panels, the 
green numbers indicate only the cores associated with molecular outflows.
Bottom: example spectra of SiO and CO lines.  The black vertical 
dashed lines represent the $v_{\rm lsr}$ of the clump. 
The directions of outflow \#1 and \#2 toward \#1--G341.03 are approximately 
along northeast-southwest and southeast-northwest, respectively. 
}
\end{figure*}

\begin{figure*}\ContinuedFloat % Example image
\center
\subfloat[]{
 \includegraphics[scale=0.5]{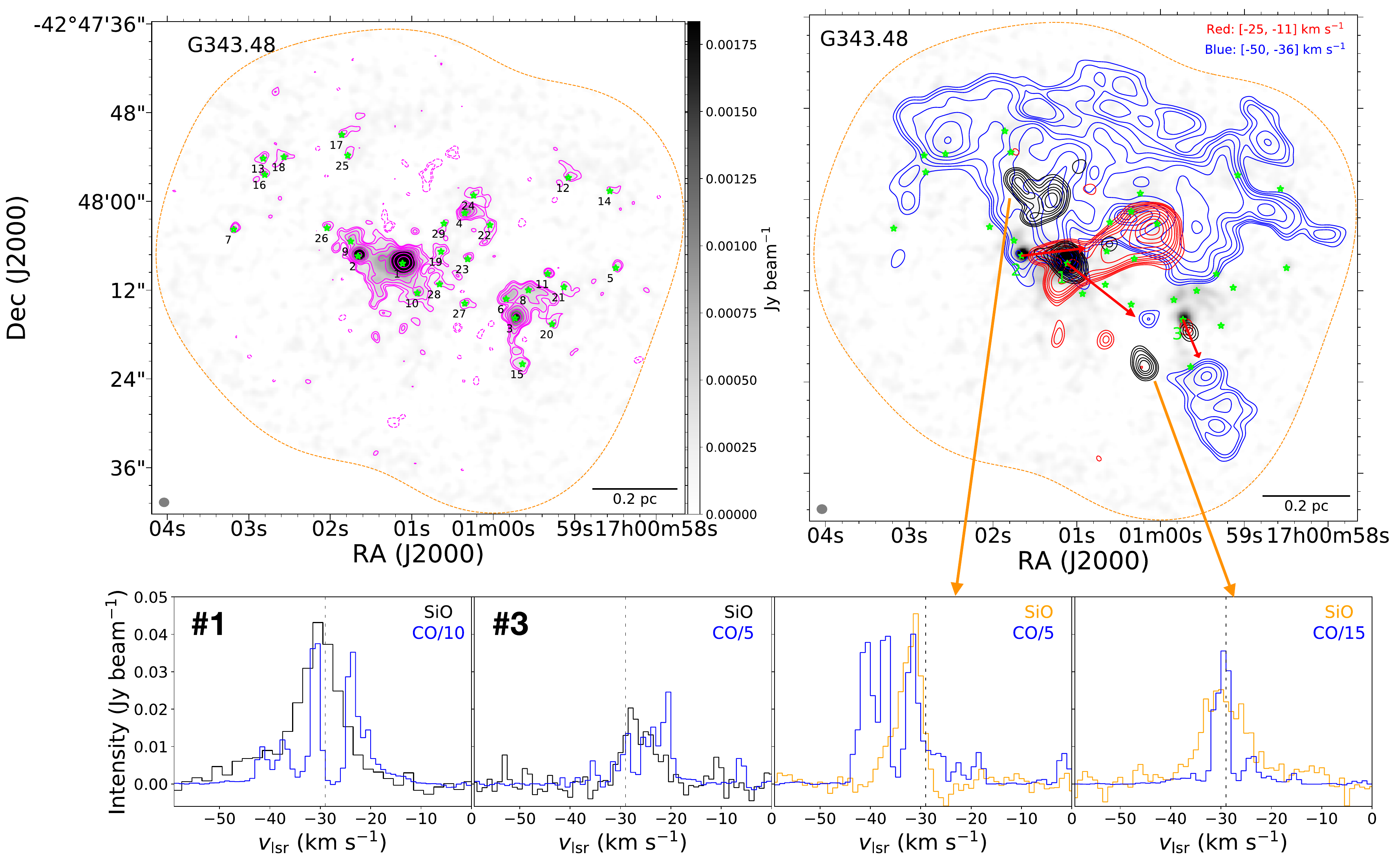}
}\\
\caption{Presented here are individual plots for Figure \ref{fig:outfmap} 
which are available as the online material. 
Dashed orange contour indicates 20\% of the sensitivity level of the 
mosaic from the ALMA continuum image.
The green stars indicate the identified dense cores. 
Upper left: ALMA 1.3~mm continuum image of G343.48. The contour levels 
are (-4, -3, 3, 4, 6, 8, 12, 20, 40, 100)$\times \sigma$, with 
$\sigma = $0.068 mJy beam$^{-1}$.  
Upper right: The blue-shifted (blue contours) and red-shifted (red contours) 
components of CO emission, and SiO velocity-integrated intensity 
(black contours, integration range is between -42 and -19 \kms)  overlaid on the 
continuum emission.  The blue and 
red contours levels are (5, 7, 9, 14, 20, 27, 35, 44, 54, 74)$\times \sigma$, 
where the $\sigma$ are 0.04 Jy beam$^{-1}$ km s$^{-1}$ and 
0.06  Jy beam$^{-1}$ km s$^{-1}$ for blue and red contours, respectively.  
The black contours are (3, 4, 5, 6, 7, 9, 11, 14, 20, 27)$\times \sigma$,
with $\sigma =$ 0.015 Jy beam$^{-1}$ km s$^{-1}$. 
Green stars show the position of the dust cores. 
On the left side panels, black numbers 
correspond to the name of all cores. On the right side panels, the 
green numbers indicate only the cores associated with molecular outflows.
Bottom: example spectra of SiO and CO lines.  The black vertical 
dashed lines represent the $v_{\rm lsr}$ of the clump. 
}
\end{figure*}
%---------------------

%\end{CJK*}
\end{document}